\documentclass[twocolumn, twocolappendix]{aastex631}
\graphicspath{{./figs/}}

\usepackage{amsmath}
\usepackage[thinc]{esdiff} 
\usepackage{xifthen}
\usepackage{txfonts}
\usepackage{xfrac}
\usepackage{enumitem}

\newcommand{\tr}[1]{\textrm{#1}}

\newcommand{\response}[1]{{#1}}

\newcommand{\tlifetime}{\tau_f}

\newcommand{\gsmfNormTot}{\Psi_0}
\newcommand{\gsmfnorm}{\psi_0}

\newcommand{\gsmfMassTot}{M_\psi}
\newcommand{\gsmfmass}{m_{\psi,0}}

\newcommand{\mmbamp}{\mu}
\newcommand{\mmbplaw}{\alpha_\mu}
\newcommand{\mmbscatter}{\epsilon_\mu}

\newcommand{\hardrchar}{a_c}
\newcommand{\hardainit}{a_\tr{init}}
\newcommand{\hardaisco}{a_\tr{isco}}
\newcommand{\hardnuinner}{\nu_\tr{inner}}
\newcommand{\hardnuouter}{\nu_\tr{outer}}
\newcommand{\harddadtnorm}{H_a}
\newcommand{\thardf}{\tau_f}

\newcommand{\mbh}{M_\tr{BH}}
\newcommand{\mbulge}{M_\tr{bulge}}
\newcommand{\mmbulge}{{\mbh\tr{--}\mbulge}}

\newcommand{\mmbulgefbulge}{f_{\star,\tr{bulge}}}
\newcommand{\gsmffunc}{\Psi}

\newcommand{\tgal}{T_\tr{gal-gal}}

\newcommand{\msol}{\tr{M}_{\odot}}

\newcommand{\pyr}{\textrm{yr}^{-1}}

\newcommand{\ndens}{\eta}
\newcommand{\ndensgalgal}{\eta_\tr{gal-gal}}

\newcommand{\mstar}{m_{\star1}}

\newcommand{\qstar}{q_\star}

\newcommand{\mchirp}{\mathcal{M}}     
\newcommand{\distcom}{d_c}   
\newcommand{\dcbg}{\langle d_c\rangle_\tr{BG}}   
\newcommand{\massbg}{\langle M\rangle_\tr{BG}}   
\newcommand{\dcss}{d_{c,\tr{SS}}}   
\newcommand{\massss}{M_\tr{SS}}   

\newcommand{\hc}{h_\tr{c}}
\newcommand{\hs}{h_\tr{s}}
\newcommand{\hscirc}{h_\tr{s,circ}}
\newcommand{\hcbg}{h_\tr{c,BG}}
\newcommand{\hcss}{h_\tr{c,SS}}

\newcommand{\snrssi}{\tr{SNR}_\tr{SS,i}}
\newcommand{\dpssi}{\tr{DP}_\tr{SS,i}}
\newcommand{\dpss}{\tr{DP}_\tr{SS}}
\newcommand{\evss}{\langle N_\tr{SS} \rangle}

\newcommand{\snrbg}{\tr{SNR}_\tr{BG}}
\newcommand{\dpbg}{\tr{DP}_\tr{BG}}

\newcommand{\redamp}{A_\tr{RN}}
\newcommand{\redgamma}{\gamma_\tr{RN}}

\newcommand{\favg}{\langle f_\tr{SS} \rangle}
\newcommand{\fstat}{\mathcal{F}_e}

\newcommand{\lr}[2][]{
    \ifthenelse{\equal{#1}{}}{
        {\left(#2\right)}
    }{
        {\left(#2\right)}^{#1}
    }
}

\newcommand{\lrs}[2][]{
    \ifthenelse{\equal{#1}{}}{
        {\left[#2\right]}
    }{
        {\left[#2\right]}^{#1}
    }
}

\newcommand{\scale}[3][]{
    \ifthenelse{\equal{#1}{}}{
        \lr{ \frac{#2}{#3} }
    }{
        {\lr[#1]{ \frac{#2}{#3} }}
    }
}

\newcommand{\figref}[1]{Fig.~\ref{#1}}

\definecolor{purple1}{rgb}{0.6, 0.0, 0.8}
\definecolor{green1}{rgb}{0.25, 0.5, 0.25}
\definecolor{red1}{rgb}{0.7, 0.15, 0.15}
\definecolor{blue1}{rgb}{0.11, 0.09, 0.71}

\shorttitle{Gravitational Wave Anisotropy and Continuous Waves from SMBHBs}
\shortauthors{Gardiner et al.}

\begin{document}


\title{Beyond the Background: Gravitational Wave Anisotropy and Continuous Waves from Supermassive Black Hole Binaries}


\author[0000-0002-8857-613X]{Emiko C. Gardiner}
\affiliation{Department of Astronomy, University of California, Berkeley 
501 Campbell Hall \#3411 
Berkeley, CA 94720, USA}

\author[0000-0002-6625-6450]{Luke Zoltan Kelley}
\affiliation{Department of Astronomy, University of California, Berkeley 
501 Campbell Hall \#3411 
Berkeley, CA 94720, USA}

\author[0009-0005-3568-3336]{Anna-Malin Lemke}
\affiliation{II. Institute of Theoretical Physics, Universität Hamburg, 22761, Hamburg, Germany}

\author[0000-0003-2898-5844]{Andrea Mitridate}
\affiliation{Deutsches Elektronen-Synchrotron DESY, Notkestr. 85, 22607 Hamburg, Germany}

\reportnum{DESY-23-132}





\begin{abstract}
Pulsar timing arrays have found evidence for a low-frequency gravitational wave background (GWB). 
Assuming the GWB is produced by supermassive black hole binaries (SMBHBs), the next gravitational wave (GW) signals astronomers anticipate are Continuous Waves (CWs) from single SMBHBs and their associated GWB anisotropy. 
The prospects for detecting CWs and anisotropy are highly dependent on the astrophysics of SMBHB populations. 
Thus, information from single sources can break degeneracies in astrophysical models and place much more stringent constraints than the GWB alone.
We simulate and evolve SMBHB populations, model their GWs, and calculate their anisotropy and detectability. 
We investigate how varying components of our semi-analytic model, including the galaxy stellar mass function, the SMBH--host galaxy relation ($\mmbulge$), and the binary evolution prescription impact the expected detections. 
The CW occurrence rate is greatest for few total binaries, high SMBHB masses, large scatter in $\mmbulge$, and long hardening times. 
The occurrence rate depends most on the binary evolution parameters, implying that CWs offer a novel avenue to probe binary evolution.
The most detectable CW sources are in the lowest frequency bin for a 16.03-year PTA, have masses from $\sim\!\!10^9-10^{10}\msol$, and are $\sim\!\!1$ Gpc away.
The level of anisotropy increases with frequency, with the angular power spectrum over multipole modes $\ell$ varying in low-frequency $C_{\ell>0}/C_0$ from $\sim\!\!5\times 10^{-3}$ to $\sim\!\!2\times10^{-1}$, depending on the model; typical values are near current upper limits. 
Observing this anisotropy would support SMBHB models for the GWB over cosmological models, which tend to be isotropic.

\end{abstract}


\color{black}
\section{Introduction} \label{sec:intro}
Supermassive black hole binaries (SMBHBs) are predicted to result from galaxy mergers. Two galaxies, each hosting a central supermassive black hole (SMBH) \citep{Richstone+1998_SMBHs}, merge as predicted by hierarchical structure formation \citep{Lacey+1993_galaxymergers}. Then, their SMBHs sink to the center of the merged galaxies via dynamical friction, become gravitationally bound, and form a binary with $\sim$pc separation. Stellar scattering and circumbinary disk torques harden the binary to small separations ($\sim\!\! 10^{-2}~\tr{pc}$) \response{\citep{BBR_1980, quinlan-1996, cuadra+2009, Kelley+2017}} beyond which they evolve primarily by emitting gravitational waves.

The superposition of these continuous waves (CWs) from many SMBHBs across the universe creates an incoherent stochastic gravitational wave background (GWB) \response{\citep{rajagopal+romani_1995, jaffe+backer_2003, burkespolaor+2019_review}}, like that for which pulsar timing arrays (PTAs) have recently found strong evidence \citep{ng+23_gwb, epta+23_gwb, ppta+23_gwb, cpta+23}. In the likely scenario that the PTA-observed GWB is produced by SMBHBs \response{\citep{ng+23_astro, epta+23_physics}}, CWs from individual, loud SMBHBs are the next highly-anticipated GW signal PTAs could detect. PTA searches have yet to find a CW source \citep{ng+23_individuals, epta+23_individuals}, but
 simulation-based predictions suggest single source CWs could be detected within a few years of the GWB \citep{Kelley+2018_ss}. 
These single sources will likely brighten certain regions of the gravitational wave sky, inducing anisotropy in the background \citep{Pol+2022} before they can be individually resolved. 

Cosmological models (cosmic inflation, phase transitions, cosmic strings, domain walls, etc.) for the GWB have also been suggested \response{\citep{ng+23_newphysics, epta+23_physics}}. These are more likely to be isotropic. Thus, measuring anisotropy in the GWB would serve as compelling evidence for SMBHBs being the source. This anisotropy has been predicted using analytic \citep{Mingarelli+2013, hotinli, satopolito+23}, semi-analytic \citep{Mingarelli+2017}, and simulation-based \citep{Taylor+Gair2013, taylor+2020, Becsy+2022, ng+23_anis} methods. 
We conduct the first study into what information content GW anisotropy contains about astrophysical models. 
Further, this paper offers the first look at how single-source detection statistics and anisotropy are related. 

Past works have predicted the amplitude and shape of the GWB using host galaxy populations generated from galaxy formation simulations \citep{Kelley+2017, Becsy+2022, Sykes+2022_gwb, becsy+2023}, dark matter (DM) merger trees \citep{Izquierdo+2022_gwb}, galaxy catalogs \citep{Mingarelli+2017}, or semi-analytic models \citep{Sesana+2008, ng+23_astro}, and others have predicted single-source CWs using galaxy simulations \citep{Kelley+2018_ss}, DM merger trees \citep{Sesana+2009_ss}, and semi-analytic SMBHB assembly models \citep{Rosado+2015}. 
Such studies have historically focused on specific hardening processes \citep{Kelley+2018_ss, Siwek+2020} or accretion scenarios and SMBH--host galaxy relations \citep{Sesana+2005}. 
To advance this field, we predict the parametric dependence of the likelihood and nature of low-frequency CW signals on the most complete SMBHB assembly \textit{and} evolution models to date. This is the first systematic investigation of model parameter space and the information content of single CW sources.

We generate SMBHB populations using \texttt{holodeck} \citep{holodeck} as explained in \S \ref{sec:meth_smbhbs}, extract the loudest single sources, and calculate gravitational waves and binary properties of both the background and single sources as described in \S \ref{sec:meth_gws}. We present the resulting characteristic strain spectra, total masses, and final comoving distances, for variations on several model components, including the galaxy stellar mass function (GSMF) (\S \ref{sec:results_gsmf}), the SMBH-host relations (\S \ref{sec:results_mmb}), and the binary evolution (\S \ref{sec:results_phenom}). Then we calculate single source detection statistics for simulated PTAs using the methods described in \S \ref{sec:meth_detstats}. The resulting single source occurrence rates and predicted properties (mass, distance, and frequency) are given in \S \ref{sec:results_ds_evss} and \S \ref{sec:results_ds_properties}, respectively. Finally, we calculate the GWB anisotropy from these SMBHB populations as described in \S \ref{sec:meth_anisotropy}, with the resulting angular power spectrum presented in \S \ref{sec:results_anisotropy}. We discuss caveats to our model and future steps in \S \ref{sec:discussion} and summarize our key findings in \S \ref{sec:conclusions}.

\color{black}

\section{Methods} \label{sec:methods}

\subsection{Model for SMBHB Populations}
\label{sec:meth_smbhbs}

Using \texttt{holodeck} \citep{holodeck} we assemble a population of galaxy mergers with comoving volumetric number density
$\ndensgalgal \equiv dN_\tr{gal-gal}/dV_c$ \citep{Chen+2019}, 
\begin{equation}
\label{eq:ndens_galgal}
        \diffp{\ndensgalgal}{{\mstar}{\qstar}{z}} = \frac{\gsmffunc(\mstar,z')}{\mstar \ln\! \lr{10}} \, \frac{P(\mstar,\qstar,z')}{\tgal(\mstar,\qstar,z')} \diffp{t}{{z'}}.
\end{equation}
We direct the reader to \citet{ng+23_astro} for a full description of the semi-analytic model components, including the galaxy pair fraction $P$ and galaxy merger time $\tgal$, both of which are power-law functions of galaxy stellar mass $\mstar$, galaxy mass ratio $\qstar$ and initial redshift $z'$.

The components of the model that we investigate in this paper are: (1) the normalization $\gsmfnorm$ and characteristic mass $\gsmfmass$ of the galaxy stellar mass function (GSMF) $\gsmffunc$, (2) the dimensionless mass normalization $\mmbamp$ and intrinsic scatter $\mmbscatter$ of the SMBH mass--bulge mass ($\mbh$--$\mbulge$) relation, and (3) the binary lifetime $\thardf$ and `inner regime' power-law index $\hardnuinner$ of the phenomenological hardening model, each of which are summarized below. 
We study the effects of each of these six parameters in isolation, by independently varying one parameter across the range listed in Table \ref{tab:varpars} while fixing the five other parameters to the fiducial values listed there and all other model components to the fiducial values in \citet{ng+23_astro} Table B1. \response{We examine a wide range of parameter space corresponding to the same range explored in \citet{ng+23_astro}. While this range extends beyond currently predicted uncertainties, several model components remain poorly constrained by observations, and exploring above and below likely values allows us to clearly identify the impact of each model component and where GWs provide constraining power.}

\begin{table}[]
    \begin{tabular}{c|c|c|c}
         Model Component & Parameter & Range & Fiducial\tablenotemark{a} \\
         \hline
            GSMF    & $\gsmfnorm$   & [-3.5, -1.5] & -2.50 \\
                    & $\gsmfmass$   & [10.5, 12.5] & 11.50 \\
         \hline
         $\mmbulge$ & $\mmbamp$     & [7.6, 9.0] & 8.30 \\
                    & $\mmbscatter$ & [0.0, 0.9] & 0.45 \\
         \hline
         phenom $\lr{\frac{da}{dt}}$    & $\thardf$     & [0.1, 11.0] & 5.55\\
                                        & $\hardnuinner$ & [-1.5, 0.0] & -0.75\\
    \end{tabular}
    \caption{Astrophysical parameters of the model components investigated in this paper, while the rest remain fixed to the fiducial values in \citet{ng+23_astro} Table B1.}
    \tablenotetext{a}{The fiducial values are calculated as the mean across the varying ranges, which correspond to the uniform priors in \citet{ng+23_astro}.}
    \label{tab:varpars}
\end{table}

\paragraph{\bf{GSMF}} 
The GSMF is the number density of galaxies per decade of stellar mass that determines the initial distribution of galaxies. We represent the GSMF as a single Schechter function \citep{Schechter-1976},
\begin{equation}
  \label{eq:gsmf_schechter}
    \gsmffunc(\mstar, z) = \ln(10)\gsmfNormTot\cdot \lrs{\frac{\mstar}{\gsmfMassTot} }^{\alpha_\psi} \exp \lr{-\frac{\mstar}{\gsmfMassTot} }, \\  
\end{equation}
where $\mstar$ is the primary galaxy stellar mass; $\gsmfNormTot$, $\gsmfMassTot$, and $\alpha_\psi$ are phenomenological functions parameterized over redshift as in \citet{Chen+2019} such that 
\begin{eqnarray}
    \label{eq:gsmf_params}
    \log_{10}\lr{\gsmfNormTot / \tr{Mpc}^{-3}} = & \, \psi_0 + \psi_z \cdot z, \nonumber \\ 
    \log_{10}\lr{\gsmfMassTot / \msol} = & \, m_{\psi,0} + m_{\psi,z} \cdot z, \\
    \alpha_\psi = & \, 1 + \alpha_{\psi,0} + \alpha_{\psi,z} \cdot z. \nonumber
\end{eqnarray}

\response{
The population of merged galaxies is derived from this distribution through the galaxy pair fraction and galaxy merger time specified in \citet{ng+23_astro} using a fixed power-law dependence on mass, redshift, and mass ratio. While the GSMF is well constrained by current astrophysics, varying our GSMF normalization captures changes in the coefficients normalizing the pair fraction and merger time as well, thus representing the overall number of \textit{merged} galaxies.}

\paragraph{\bf{SMBH--Host Relation}}
The SMBH masses are related to their host galaxies' bulge masses $\mbulge$ by assuming an $\mbh$--$\mbulge$ relation defined by dimensionless mass normalization $\log_{10} \mu$ and power-law index $\mmbplaw$, in addition to a random normal distribution of $\log_{10}$ scatter $\mathcal{N}\lr{0, \mmbscatter}$ with standard deviation $\mmbscatter$:
\begin{equation}
\label{eq:mmbulge_relation}
\log_{10}\! \lr{\mbh/\msol} = \mmbamp + \mmbplaw \log_{10}\!\scale{\mbulge}{10^{11} \, \msol} + \mathcal{N}\lr{0, \mmbscatter}.    
\end{equation}
The bulge mass is calculated as a fraction of the total galaxy stellar mass $= \mmbulgefbulge \cdot \mstar$, with $\mmbulgefbulge=0.615$ based on empirical observations from \citet{Lang+2014} and \citet{Bluck+2014}.

Applying the $\mbh$--$\mbulge$ relation in Eq. \ref{eq:mmbulge_relation}, the number density of merged galaxies in Eq. \ref{eq:ndens_galgal} translates to the number density of SMBHBs by
\begin{equation}
        \diffp{\ndens}{{M}{q}{z}} = \diffp{\ndensgalgal}{{\mstar}{\qstar}{z}} \: \diffp{\mstar}{M} \: \diffp{\qstar}{q}.
\end{equation}

\paragraph{\bf{Hardening}} To model gravitational waves detectable by PTAs, the population of SMBHBs must be evolved in time and separation to rest frame orbital frequencies $f_p$ corresponding to GW frequencies of a few nHz. This binary hardening is described in terms of a rate of decreasing separation, $da/dt = (da/dt)_\tr{gw} + (da/dt)_\tr{phenom}$, i.e.~the sum of a GW component 
\begin{equation}
    \label{eq:gw_hard}
    \frac{d a}{d t}\bigg|_\tr{gw} = -\frac{64 \, G^3}{5 \, c^5} \frac{m_1 \, m_2 \, M}{a^3}, 
\end{equation}
and a phenomenological component, 
\begin{equation}
\label{eq:hard_phenom}
    \frac{d a}{d t}\bigg|_\tr{phenom} = \harddadtnorm \cdot \lr[1-\hardnuinner]{\frac{a}{\hardrchar}} \cdot \lr[\hardnuinner-\hardnuouter]{1 + \frac{a}{\hardrchar}}.
\end{equation}
A double power law allows for distinct asymptotic behavior in the small-separation `inner' regime and large-separation `outer' regime, distinguished by a critical break separation $a_c$. $\harddadtnorm$ is a normalization factor, calibrated for every binary such that it has a total binary lifetime from initial separation $\hardainit$ to coalescence at the innermost stable circular orbit $\hardaisco$ of
\begin{equation}
    \tlifetime = \int_{\hardainit}^{\hardaisco} \scale[-1]{da}{dt} da.
\end{equation}
This serves as a self-consistent approach to modeling binary evolution, without depending on assumptions about the binary hardening processes or galactic environment. We also investigate the effects of varying our four GSMF and $\mbh$--$\mbulge$ parameters for the GW-only model as in \citet{ng+23_astro}, which is not self-consistent because GWs alone cannot bring the binaries to small enough separations to emit nHz GWs, but serves as a useful comparison.

\subsection{Binary Properties and Gravitational Waves}
\label{sec:meth_gws}
The analytic model described in \S \ref{sec:meth_smbhbs} determines a comoving volumetric number density of SMBHBs $\frac{\partial^3 \eta}{\partial M \partial q \partial z}$, from which we calculate a continuous number of SMBHBs per mass $M$, ratio $q$, redshift $z$ (at the time of GW emission), and log rest-frame orbital frequency $\ln f_p$ \citep{Sesana+2008}: 
\begin{equation}
    \label{eq:number_density_to_number_frequency}
    \diffp{N}{{M} {q} {z} {\ln f_p}} = \diffp{\ndens}{{M} {q} {z}} \diffp{t}{{\ln f_p}} \diffp{z}{{t}} \diffp{V_c}{{z}}. 
\end{equation}
This continuous distribution sets a fractional expectation value for the number of binaries. In reality, gravitational waves are produced by a discrete population of binaries. We generate random universe realizations of this population by selecting a number of binaries $N(M,q,z,f)$ in each parameter bin of $\Delta M$, $\Delta q$, $\Delta z$, and $\Delta (\ln f_p)$ from a Poisson distribution ($\mathcal{P}$) centered at the aforementioned expectation value, 
\begin{eqnarray}
    \label{eq:number_sampling}
    N(M,\!q,\!z,\!f) = \quad \quad \quad \quad \quad & \nonumber \\
     \mathcal{P}\,\Big( \diffp{N}{{M'} {q'} {z'}{\!\ln f_p'}} &
    \Delta M' \! \Delta q'\!\Delta z' \! \Delta \!\ln f'\Big)|_{M, q, z, f_p}
\end{eqnarray}
We assume circular orbits for all binaries and assign them the $M$, $q$, $z$, and $f_p$ values corresponding to their bin centers. These define their chirp mass $\nobreak{\mchirp \equiv \left(m_1 m_2\right)^{3/5} / M^{1/5} = M q^{3/5} / \left(1 + q\right)^{6/5}}$
comoving distance $d_c$, 
observer-frame GW frequency $\nobreak{f=(2 f_p)/(1+z)}$,
and (sky and polarization averaged) GW strain amplitude of \citep{Finn+2000}
\begin{equation}
    \label{eq:strain_amp}
    \hscirc^2(f_p) = \frac{32}{5 c^8} \,  \frac{\left(G\mchirp\right)^{10/3}}{\distcom^2} \lr[4/3]{2\pi f_p}.
\end{equation}

Because the loudest single source may not be the most detectable, depending on sky position, inclination, etc., the \textit{ten} loudest single sources (SS, i.e.~with the greatest $\hs$) in each frequency bin are then extracted from this population. Their individual characteristic strains are calculated as 
\citep{Rosado+2015}
\begin{equation}
\label{eq:char_strain_ss}
    \hcss^2(f) =  \hs^2(f_p) \scale{f}{\Delta f} .
\end{equation}
Here, $\Delta f$ is the frequency bin width and arises when considering a finite number of sources in finite frequency bins $N\sim f*T\sim f/\Delta f$, over an observing duration $T$.

The GWB is then calculated as the sum of gravitational waves from all background (BG) binaries 
\begin{equation}
\label{eq:char_strain_bg}
    \hcbg^2(f)\!=\!\!\!\sum_{M,q,z,f}\!\!\, N(M,q,z,f) \hs^2(f_p) \frac{ f}{\Delta\!\ f}.
\end{equation}
Here, we define BG binaries to include all but the single loudest at each frequency because the most immediate observational application of this work will be the detection of \textit{one} CW source, before PTAs can resolve multiple of them. 
When considering the GW-only model without phenomenological hardening, this in combination with the GW hardening rate $(da/dt)_\tr{gw}$, leads to the $\hcbg \propto f^{-2/3}$ power law often used as a comparison point for the characteristic strain spectra. In reality, we expect deviations from this power law not only due to the phenomenological hardening before GWs dominate the evolution \citep{kocsis+2011}, but also due to the discretization of sources where the power-law would otherwise predict fractional binaries \citep{Sesana+2008}.
\response{We also calculate the characteristic mass, mass ratio, redshift, comoving distance, separation, and angular separation for the SMBHBs contributing most to the background at each frequency. To do so, we perform  an $\hcbg$-weighted average over all BG binaries emitting at that frequency.}

\subsection{Detection Statistics}
\label{sec:meth_detstats}
Given the $\hcss$ and $\hcbg$ spectra, we calculate SS and BG detection statistics following the formalism in \citet{Rosado+2015}. This includes the background signal-to-noise ratio ($\snrbg$), and detection probability ($\dpbg$), and each individual source's SNR ($\snrssi$) and detection probability ($\dpssi$). The probability of detecting \textit{any} single source is then  \citep{Rosado+2015}
\begin{equation}
\label{eq:dpss}
  \dpss = 1 - \prod_i [1-\dpssi]  
\end{equation}
and the expected number of detections for that realization is 
\begin{equation}
\label{eq:evss}
\evss = \sum_i \dpssi    .
\end{equation}

In this prescription, single source detection probabilities are given by integrating over the $\fstat$-statistic from some threshold $\bar{\mathcal{F}}_e$ to infinity, where $\mathcal{\bar{F}}_e$ is set to give a false alarm probability (FAP) of $10^{-3}$. Even with no signal present, the area under this curve will produce a nonzero total detection probability ($\dpss$) equal to the FAP. Thus, $10^{-3}$ is the lower limit on detection probabilities calculated in Eq.\ \ref{eq:dpss} and should be treated effectively as 0. 

In light of the strong evidence for a GWB in current PTA data \citep{ng+23_gwb, epta+23_gwb, ppta+23_gwb, cpta+23}, we study the single source detection probability under the same conditions that are likely to produce measurable GWB evidence by calibrating every realization to $\dpbg=0.50$. We use a white-noise-only simulated PTA of 40 pulsars at randomly assigned sky positions, 16.03 yr duration \citep[corresponding to][]{ng+23_gwb}, and 0.20 yr cadence $\Delta t$. Our fiducial method of calibration is to vary the level of white noise $S_\tr{WN}$, given by the error in pulsar times of arrival (ToAs) $\sigma$ \citep{Rosado+2015}:
\begin{equation}
\label{eq:white_noise}
    S_\tr{WN} = 2 \Delta t \sigma^2,
\end{equation} 
until achieving $0.49<\dpbg<0.51$. We calculate $\evss$ using the same pulsar positions and $\sigma$, with all characteristic strains except that of the source in question considered additional noise,
\begin{equation}
\label{eq:rest_noise}
    S_{\tr{rest},i} = \frac{\hcbg^2 + \sum_{j\neq i} h_{c,\tr{SS},j}^2}{12\pi^2 f^3}.
\end{equation}
Then we normalize for small variations around $\dpbg=0.50$ with
\begin{equation}
\label{eq:evss_norm}
    \evss[\dpbg=0.50] = \frac{\evss[\dpbg\approx0.50] }{\dpbg} \times 0.50
\end{equation}
For one realization of BG and SS characteristic strain, we calibrate the PTA to the background, then create 100 `sky realizations'--the random position, inclination, polarization, and phase assignment for single sources--and conduct SS detection statistics for each. By calculating detection statistics for 10 single sources for each frequency of each realization, we allow for the most detectable to depend on both strain amplitude and random location/orientation. This is repeated for 500  `strain realizations' of $\hcbg$ and $\hcss$--those created by Poisson sampling in Eq.\ \ref{eq:number_sampling}, each with their own BG-calibrated PTA--to create 50,000 combined `strain+sky realizations.' 

Next, we predict the most likely frequencies of detection by calculating the $\dpssi$-weighted average frequency of all $n$ loudest single sources across all realizations of a given model:
\begin{equation}
\label{eq:f_avg}  
\favg = \frac{\sum_i \dpssi f_{\tr{SS},i}}{\sum_i \dpssi}
\end{equation}
with weighted standard deviation:
\begin{equation}
\label{eq:f_std}
\sigma_{\favg} = \frac{\sum_i \dpssi (f_{\tr{SS},i} - \favg)^2}{\frac{n-1}{n}
\sum_i \dpssi}
\end{equation}
The likely frequency of detection is sensitive to the shape of the PTA noise. Thus, we explore different noise models inspired by realistic sensitivity curves and intrinsic pulsar red noise in \S \ref{sec:app_rednoise}.

\subsection{GWB Anisotropy}
\label{sec:meth_anisotropy}
We measure the anisotropy corresponding to each model by decomposing a simulated GW sky into spherical harmonics, as in \citet{ng+23_anis}. To generate this GW sky, we create a HEALpix map \citep{healpix} of $\hc^2/\Delta \Omega$ ($\Omega$ being solid angle, or equivalently, pixel area) at each frequency of each realization by assigning the single sources to random pixels and distributing the remaining $\hcbg$ evenly among all the pixels.

Then, we can decompose this sky into an angular power spectrum of multipole modes $\ell$ and $m$ each accompanied by a coefficient $a_{\ell m}$ such that the total GW power is the sum of each $a_{\ell m}$ times the real-valued spherical harmonic $Y_{\ell m}$. The \texttt{anafast} code, via \texttt{healpy} \citep{healpy} calculates these coefficients with the estimator \citep{healpix_primer}
\begin{equation}
    a_{\ell m} = \frac{4\pi}{N_\tr{pix}} \sum_{p=0}^{N_\tr{pix}-1} Y^*_{\ell m} (\gamma_p) f(\gamma_p),
\end{equation}
where $N_\tr{pix}=12N_\tr{side}^2$ is the number of pixels, indexed by $p$ at positions $\gamma_p$, and $f(\gamma_p)$ is $\hc^2/\Delta \Omega$ in each pixel.

Using \texttt{anafast}, we calculate the corresponding angular power spectrum
\begin{equation}
    C_\ell = \frac{1}{2\ell + 1} \sum_{m=-\ell}^\ell |a_{\ell m} |^2
\end{equation} 
where $C_\ell$ represents the measure of fluctuations (i.e.~anisotropy) on the angular scale $\theta \approx 180\deg / \ell$. $C_0$ represents the purely isotropic average component, thus we normalize our results by $C_\ell / C_0$. 

This method is tested for 10, 100, and 1000 loudest single sources per frequency bin, by which point our results are insensitive to the addition of more single sources. By placing these sources randomly and treating the remaining signal as purely isotropic, we do not weigh in possible correlations with large-scale-structure, making this a conservative estimate for anisotropy. 
We also test the resolution and find $C_\ell$ to be indistinguishable for $N_\mathrm{side}=8$ up to  $N_\mathrm{side}=32$. Thus, we adopt $N_\mathrm{side}=8$ ($N_\tr{pix}=768$) to efficiently calculate the spherical harmonic decomposition for each realization and present the results in \S \ref{sec:results_anisotropy}.

\color{black}



\section{Results} \label{sec:results}

\subsection{Characteristic Strain and Binary Properties} \label{sec:results_hcprop}

In this section we present the characteristic strain $\hc$, total mass $M$, and final comoving distance $\distcom$ of GWB and CW sources as a function of frequency. 
The first column of \figref{fig:spectra_gsmf} includes three models with varying GSMF normalization: $\gsmfnorm = -3.5$, $-2.5$, and $-1.5$, while all other parameters remain fixed to their fiducial values listed in Table \ref{tab:varpars}. 
Information about the CW sources is shown in green, for these three models, respectively. 
This includes the 68\% confidence intervals (CIs, shaded regions) across 500 realizations of the single loudest source at each frequency. The 95th percentile of these sources' $\hcss$ and $\massss$, and the 5th percentile of these sources' $\dcss$ is also shown (points). 
For comparison, $\hcbg$ and the $\hcbg$-weighted average properties ($\massbg$ and $\dcbg$) of the background (all but the loudest single sources at each frequency) are shown in corresponding shades of grey, with dashed lines representing their medians. 

The same is shown for models of varying $\gsmfmass$ in the second column of \figref{fig:spectra_gsmf}, for $\mmbamp$ and $\mmbscatter$ in \figref{fig:spectra_mmb}, and for $\thardf$ and $\hardnuinner$ in \figref{fig:spectra_hard}. The following three sections describe the physical scenarios producing these results for each model component: the GSMF (\S \ref{sec:results_gsmf}), the $\mmbulge$ relation (\S \ref{sec:results_mmb}), and binary evolution (\S \ref{sec:results_phenom}).

\color{black}

\subsubsection{GSMF}\label{sec:results_gsmf}
\begin{figure}
    \centering
    \includegraphics[width=1.0\columnwidth]{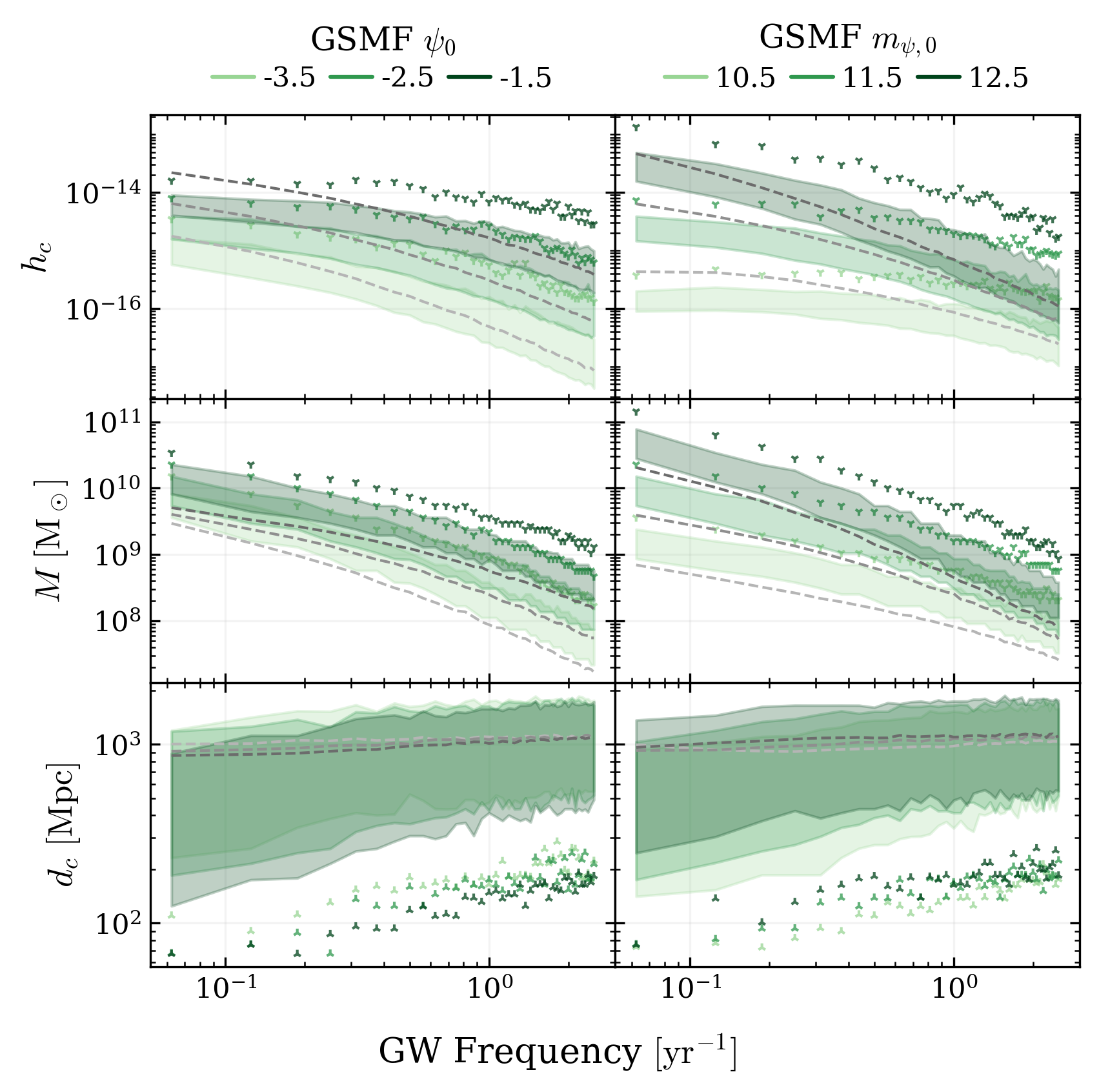}
    \caption{Characteristic strain (top row), total mass (middle row), and final comoving distance (bottom row), for varying $\gsmfnorm$ (left column) and $\gsmfmass$ (right column). Shaded green regions represent 68\% CI of the single loudest source at each frequency, with markers to indicate the 95th percentiles for $\hcss$ and $\massss$ and 5th percentiles for $\dcss$. Dashed lines represent the median background (all but the single loudest source per frequency) characteristic strain $\hcbg$ and the $\hc$-weighted background properties ($\massbg$ and $\dcbg$). $\gsmfmass$ increases from -3.5 to -2.5 to -1.5 and $\gsmfnorm$ increases from 10.5 to 11.5 to 12.5 for darkening shades of green/grey.}
    \label{fig:spectra_gsmf}
\end{figure}

The GSMF parameters shape the masses of galaxies, and thus their residing SMBHBs for fixed $\mmbulge$. 
The masses in each SMBHB directly determine its strain amplitude as $\hs \propto \mchirp^{5/3}$ (Eq.\ \ref{eq:strain_amp}). 
\response{
When considering the characteristic strain from many background sources or from a single source sampled probabilistically from a distribution, $\hcss$ and $\hcbg$ also have mass dependence in their GW strains, their hardening timescales, and their number density, such that $\hc \propto \mchirp^{5/6} \sqrt{\ndens(M, q, z) }$. 
By varying GSMF parameters, we examine the $\ndens(M,q,z)$ component of the mass distribution. 
}
Changes to the distribution of SMBHB masses also result secondarily in small $\distcom$ variations due to the mass dependence of hardening rate versus frequency (described below). 

Following the Schechter GSMF, the number of SMBHBs decreases with increasing mass. When this expectation value approaches zero, few random realizations contain a source in that bin. When $\gsmfnorm$ increases, the number of sources in every bin increases. After sampling, this translates to the loudest randomly realized sources having higher masses. 

The background sees a similar increase in the mass of its dominating sources. In addition to this, it also has a larger number of contributing sources in every mass and frequency bin. Thus, $\hcbg$ increases near-uniformly across all frequencies. Its amplification matches that of $\hcss$ at high frequencies and exceeds that of $\hcss$ at low frequencies. There, SMBHB numbers are the largest because sources harden more slowly at low frequencies. 
This is particularly true of high-mass sources, whose numbers dwindle at high frequencies where they harden quickly by emitting more GWs. 
Thus, scaling up the number of sources in every bin leads to many more massive sources contributing to the low-frequency GWB. 

The GSMF characteristic mass $\gsmfmass$ (\figref{fig:spectra_gsmf}, right-hand side), sets where the expected number of binaries drops off. Thus, varying $\gsmfmass$ only significantly impacts mass bins corresponding to the lowest end of our varying $\gsmfmass$ range ($\gsmfmass=10.5$) and above. 
Thereby, \figref{fig:spectra_gsmf} shows that increasing $\gsmfmass$ from 10.5 to 12.5 raises the $\massss$ 68\% CI most dramatically 
($\sim\! 1.6~\tr{dex}$) at low frequencies, where massive sources are more common, and more moderately ($\sim\! 0.5~\tr{dex}$) at high frequencies. 

The background also sees large increases in $\massbg$ at low frequencies 
because many loud binaries remain there, even after the loudest has been extracted, and little change in typical mass at high frequencies
where few high-mass sources remain after single-source extraction. 
This amounts to $\hcss$ having an overall more significant increase than $\hcbg$. 

\response{
While high mass sources dwindle at high frequencies due to fast GW hardening, lower mass sources are driven more quickly through the low frequency regime (for fixed binary lifetime) by environmental processes. Thus, a lower $\gsmfmass$ flattens both $\hcss$ and $\hcbg$ at low PTA-band frequencies because the loss of the most massive sources is exacerbated by the environmentally driven hardening of lower mass sources. The effects of environmental hardening (represented by $\hardnuinner$) are explored further in \S \ref{sec:results_phenom}, but here we find that when the characteristic strain is dominated by less massive binaries, environmental hardening flattens the low-frequency spectrum more. 
}

Regardless of the model, low-frequency sources tend to be nearer. 
This is because more massive sources (which are numerous at low frequencies) take longer to evolve to the PTA band, as explained in greater detail in \S \ref{sec:results_phenom}. 
Longer evolution times let these massive sources reach the PTA band at smaller redshifts and closer distances. 

\subsubsection{$\mmbulge$ Relation} \label{sec:results_mmb}
\begin{figure}
    \centering
    \includegraphics[width=1.0\columnwidth]{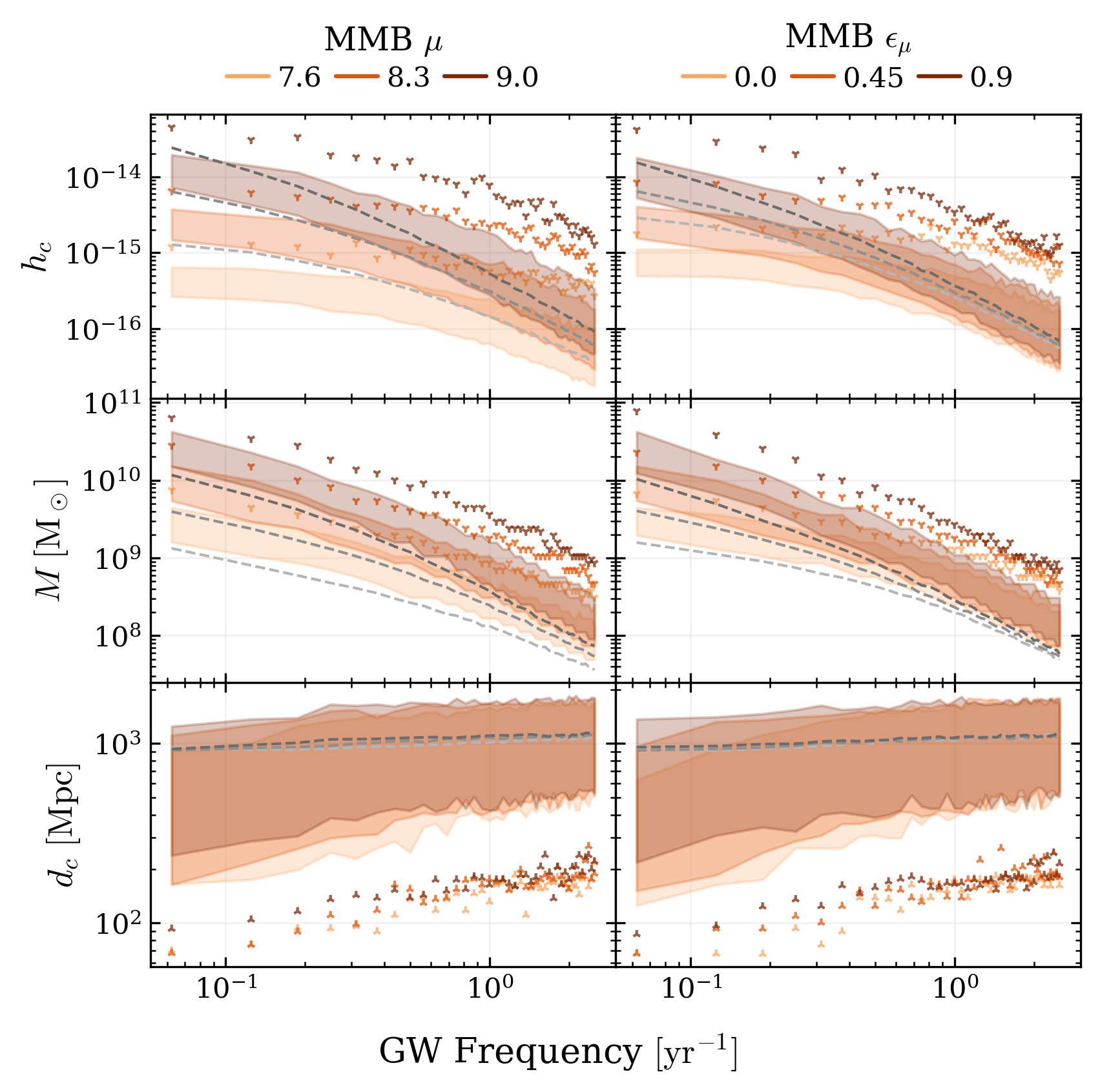}
    \caption{Same as \figref{fig:spectra_gsmf} but for the $\mmbulge$ parameters: increasing $\mmbamp$ (left column) from 7.60 to 8.30 to 9.00 and increasing $\mmbscatter$ (right column) from 0.00 to 0.45 to 0.90 for darkening shades of orange (single sources) and grey (background).}
    \label{fig:spectra_mmb}
\end{figure}
\figref{fig:spectra_mmb} shows $\hc$, $M$, and $\distcom$ for varying $\mmbulge$ parameters.
Given a population of host galaxies, the $\mmbulge$ relation in Eq.\ \ref{eq:mmbulge_relation} sets the masses of these galaxies' central SMBHBs. 
Increasing the relation's mass normalization ($\mmbamp$) shifts all SMBHBs to higher masses. 
This has negligible impact on low mass bins, where the number density is large and changes gradually with mass. 
However, at high masses, the number density drops off quickly with increasing mass. This means that shifting the expected number of sources of one bin to the next bin of increasing mass significantly raises the chances of realizing a source in that higher mass bin. Overall, this increases the odds of randomly sampling a source in any high-mass bin. 

The left column of \figref{fig:spectra_mmb} shows that a $\sim\!\!1~\tr{dex}$ increase in $\massss$ follows the increase in $\mmbamp$ from 7.6 to 9.0 at low frequencies, where massive sources are numerous.
\response{The $\sim\!\!1~\tr{dex}$  increase in mass is not as great as the $\sim\!\!1.6~\tr{dex}$ increase in mass normalization because these higher mass binaries also evolve faster.
}
At high frequencies, the 68\% CIs see a more modest increase, which follows from the fact that these represent lower mass sources.
The changes in $\hcss$ are approximately proportional to $\massss^{5/3}$ (see Eqs.\ \ref{eq:strain_amp} and \ref{eq:char_strain_ss}) with deviations below this relation owing to unequal mass ratios. 
Meanwhile, at high frequencies, the background is minimally affected by $\mmbamp$ due to the lack of massive sources, especially after the loudest have been removed. 
Ultimately, across all frequencies, increasing $\mmbamp$ raises $\hcss$ more than $\hcbg$.

Increasing scatter $\mmbscatter$ in the $\mmbulge$ relation preferentially scatters sources to higher mass bins through Eddington Bias. 
Like $\mmbamp$, this increases the odds of sampling sources from the highest mass bins. 
The right column of \figref{fig:spectra_mmb} shows that an increase in $\mmbscatter$ from 0.0 to 0.9 increases all low-frequency $\hc$, $\hcss$ slightly more than $\hcbg$. 
This preferential scattering effect becomes negligible at lower masses where the mass function flattens and number densities become large; thus, we see the $\massss$ 68\% CI and $\massbg$ medians both converge at high frequencies where lower mass sources dominate. 
\response{Similarly to the $\gsmfmass$ scenario, anytime low-mass sources are a greater contributor to the overall characteristic strain spectrum, the low-frequency end of the spectrum is also flatter, due to low-mass sources' hardening being sped up by environmental processes.}

In addition to the preferential scattering systematically increasing masses, introducing scatter to the $\mmbulge$ relation adds a second element of randomness beyond the Poisson bin sampling, widening the variance of random realizations. This is apparent as a slight widening of the 68\% CI $\massss$ and $\hcss$ regions and is more obvious in the SS 95th percentiles, which generally increase by $\gtrsim0.5$ dex, even at high frequencies where the 68\% CIs converge.

The $\mmbulge$ relation slightly impacts the distribution of $\dcss$ at low frequencies, with the 68\% CI including more distant sources when $\mmbscatter$ is higher. When scatter is low and arbitrarily high mass sources are less likely to be sampled, nearby sources tend to be the loudest. When the scatter is increased, the loudest source could instead be more massive, but further away. This effect occurs only at low frequencies because that is where $\mmbscatter$ causes a measurable change in $\massss$ and $\hcss$. We also see a frequency dependence of $\dcss$ matching that described in the last section: low-frequency single sources tend to be nearer because these sources are more massive, and thus take longer to reach the PTA band.

\subsubsection{Hardening}\label{sec:results_phenom}
\begin{figure}
    \centering
    \includegraphics[width=1.0\columnwidth]{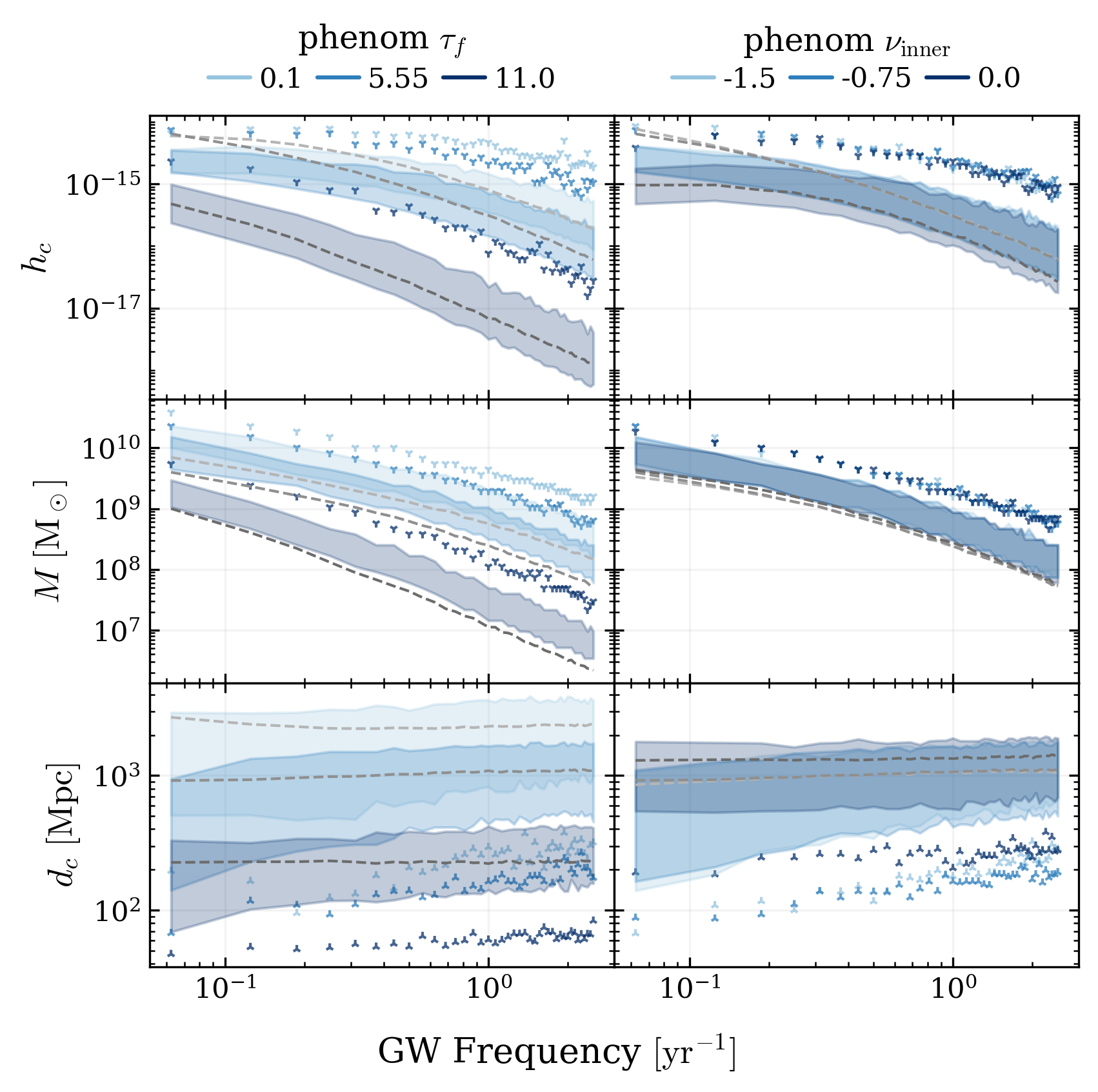}
    \caption{
    Same as \figref{fig:spectra_gsmf} but for the phenomenological hardening parameters: increasing $\thardf$ (left column) from 0.10 Gyr to 5.55 Gyr to 11.00 Gyr and flattening $\hardnuinner$ (right column) from -1.50 to -0.75 to 0.00 for darkening shades of blue (single sources) and grey (background).}
    \label{fig:spectra_hard}
\end{figure}
In our phenomenological hardening model, all sources take the same total time from galaxy merger to SMBH coalescence, set by $\thardf$. 
The way this hardening rate is distributed across the binary's lifetime depends on its mass, such that higher mass sources spend less time at high frequencies. This is because 1) more massive sources evolve more rapidly by GWs when they reach small separations and 2) more massive sources merge at larger separations. 
These combined effects require high-mass sources to evolve slower than low-mass sources at low frequencies, in order to meet the same fixed $\thardf$. 
Thus, in our model, massive sources take longer to reach the PTA frequency band. 
Lower-mass sources will reach the PTA band more quickly, and then dwell there longer as they harden slower than their high-mass counterparts until they eventually coalesce. 
The slower evolution of massive binaries to PTA frequencies means they tend to be nearer when they emit in the PTA band. 
However, those that start at too low of redshifts will not reach separations small enough for PTA-band emissions by redshift zero.

When $\thardf$ is increased, this extends the evolution time of binaries at all masses, moving the entire populations to smaller final comoving distances. 
The left column of \figref{fig:spectra_hard} shows this to be similarly true for both the background and single sources, with $\dcbg$ decreasing by $\sim\!\!1.0 ~\tr{dex}$ and $\dcss$ by $\sim\!\!1.1 ~\tr{dex}$ when $\thardf$ is raised from 0.1 Gyr (light blue) to 11.0 Gyr (dark blue). 
Secondly, when the hardening time is extended, the most massive sources are unlikely to reach small enough separations to emit at PTA-detectable frequencies, producing a decrease in mass. 
This effect is largest at high frequencies, to which binaries take the longest to evolve. 
\response{Thirdly, when the overall binary lifetime is short, even high mass binaries can be driven by non-GW hardening represented by $\hardnuinner$, causing a low-frequency flattening or turnover in the $\hc$ spectrum as described in the $\hardnuinner$ analysis below.}
The characteristic strain has a greater proportional dependence on $\massss$ than $\dcss$, so the changes in $\hcss$ with varying $\thardf$ follow the changes in $\massss$.
When considering the background, the filtration of massive sources applies to a larger \textit{number} of sources at low frequencies, where massive sources are numerous. Thus, long $\thardf$ decreases $\hcbg$ slightly more than it decreases $\hcss$ at low frequencies.

Per Eq.\ \ref{eq:hard_phenom}, $\hardnuinner$ sets the hardening rate in the small-separation regime, with the asymptotic behavior of 
\begin{equation}
    \frac{dt}{d \ln a}(a \ll \hardrchar) \sim a^{\hardnuinner}.
\end{equation}
A flatter (less negative) $\hardnuinner$ increases the hardening rate at the lowest end of the PTA frequency regime before $(d a/d t)|_\tr{gw}$ dominates \citep[see Fig.~3 in][]{ng+23_astro}. 
This represents faster hardening by processes like stellar scattering and circumbinary disk torques and produces the low-frequency turnover in both $\hcbg$ and $\hcss$ apparent in the top panels of \figref{fig:spectra_hard}. 

Single sources are only impacted at low frequencies because there, when $\hardnuinner=0$, even the most massive sources will have phenomenological hardening dominate GW hardening. \response{The same is true of massive binaries when the total binary lifetime is short.} 
Lower mass binaries can be dominated by phenomenological hardening up to higher frequencies as their GW emission is weaker. Thus, the background--including contributions from lower mass binaries--sees a lower $\hcbg$ across all frequencies for flat $\hardnuinner$.

The bottom right panel of \figref{fig:spectra_hard} shows the 68\% CI of low-frequency single sources to be more distant for flat $\hardnuinner$ than either of the other two cases. 
We attribute this to a selection bias: when $\hardnuinner$ is flat, massive sources evolve through the low end of the PTA band quickly. Thus, there are fewer of them. While any individual source is just as able and likely to reach small distances, having fewer of them decreases the odds of having an especially close source. 
These trends only appear in the 68\% CI because the lower bounds of $\dcss$ for $\hardnuinner$ represent the less-common cases where one of the few loud sources just so happens to be nearby. 
This source can be as near under flat $\hardnuinner$ as it can for steep $\hardnuinner$; it is just less likely to exist in the first place. 

\color{black}
\subsection{Astrophysical Dependence of CW Detections} \label{sec:results_ds}

\subsubsection{CW Detection Occurrence Rate} \label{sec:results_ds_evss}

\begin{figure}
    \centering
    \includegraphics[width=\columnwidth]{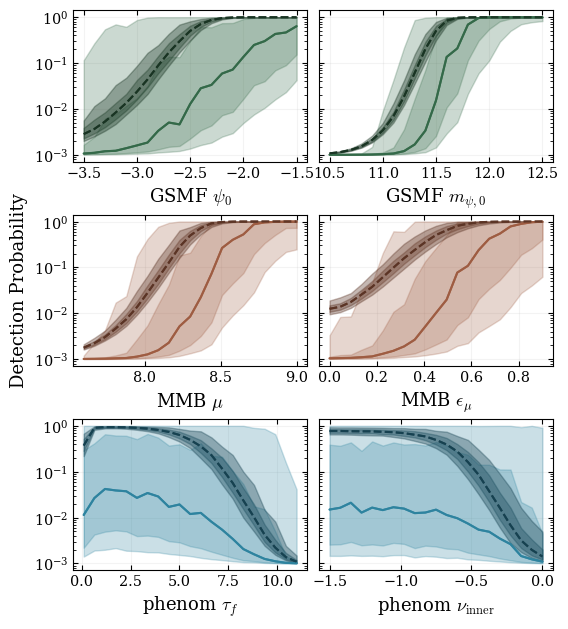}
    \caption{Detection probability for a PTA calibrated to the median $\dpbg$ of the mean parameter model (i.e.~using the fiducial values in Table \ref{tab:varpars}). 
    The PTA is calibrated once for each panel's mean phenomenological model, and once for each panel's mean GW-only model. 
    The single source detection probability is shown in color, for varying GSMF parameters in green, varying $\mmbulge$ parameters in orange, and varying hardening parameters in blue. 
    The $\dpss$ medians are represented by solid lines and the 68\% and 95\% CIs are shaded. 
    The background detection probability ($\dpbg$) is given in darker shades of the same colors, with medians as dashed lines and 68\% and 95\% CI shaded. 
    }
    \label{fig:dpboth_wn}
\end{figure}
\begin{figure}
    \centering
    \includegraphics[width=\columnwidth]{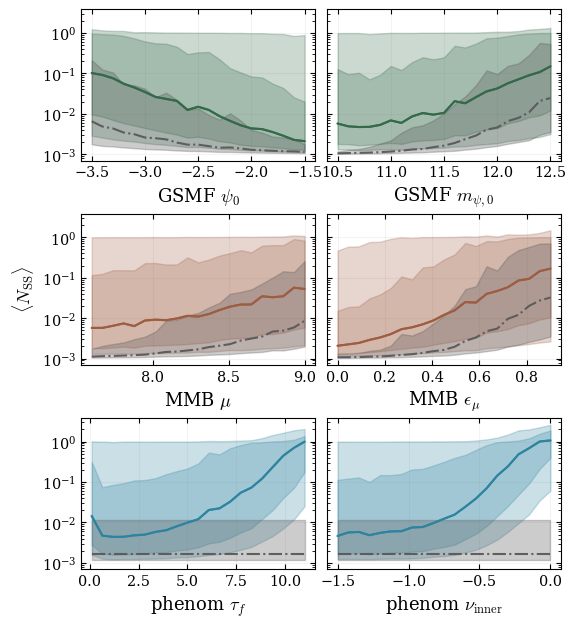}
    \caption{Expected number of single-source detections $\evss$ for a white-noise-only PTA calibrated independently to a 50\% background detection probability for each parameter and realization. $\evss$ is given as a function of varying GSMF parameters (\textit{top row}) in green, $\mmbulge$ parameters (\textit{middle row}) in orange, and phenomenological hardening parameters (\textit{bottom row} in blue) for the phenomenological hardening model. The medians are solid lines and the 68\% and 95\% CI are shaded. $\evss$ for the GW-only hardening model has medians represented by dash-dotted lines and 68\% CI shaded in grey. These are replaced in the bottom row by constant values corresponding to the fiducial GSMF and $\mmbulge$ model because with GW-only hardening there are no $\thardf$ and $\hardnuinner$ to vary. 
    }
    \label{fig:evss_wn}
\end{figure}

\figref{fig:dpboth_wn} shows the single source detection probability ($\dpss$) and the background detection probability ($\dpbg$) as a function of each varying parameter, for a fixed-PTA configuration. This `fixed-PTA' method involves calibrating the PTA's noise level so that the median $\hcbg$ across all realizations of the mean-parameter model (i.e.~with the fiducial parameter values listed in Table \ref{tab:varpars}) has a 50\% $\dpbg$. 
For example, the top left panel represents varying $\gsmfnorm$, so the PTA is calibrated to the median $\hcbg$ across 500 realizations of the $\gsmfnorm=-2.5$ model. This PTA is used throughout the rest of the varying $\gsmfnorm$ with phenomenological hardening analysis. 
The resulting $\dpss$ medians are represented by a solid green line, as well as 50\% and 95\% $\dpss$ CI as green-shaded regions. The resulting $\dpbg$ medians are represented by a dashed darker green line, with 50\% and 95\% CI as darker green shaded regions. The rest of the panels show the same, but for varying $\gsmfmass$, $\mmbamp$, $\mmbscatter$, $\thardf$, and $\hardnuinner$ as labeled.

In all cases, the $\dpss$ medians remain below $\dpbg$, consistent with the expectation for GWB detection to occur before CW detection \citep[e.g.][]{Rosado+2015}. $\dpbg$ is remarkably well constrained (95\% CI spanning $\lesssim \!\! 0.5~\tr{dex}$), while $\dpss$ 95\% CI often range all the way from $\sim\!\! 10^{-3}$ to $\sim\!\!10^0$. These 95\% CIs of $\dpss$ can exceed $\dpbg$ in a few corners of parameter space, most notably for $\gsmfnorm\lesssim-2.3$, $\thardf\gtrsim5 ~\tr{Gyr}$, and $\hardnuinner \gtrsim -0.75$. Thus, low $\gsmfmass$, long $\thardf$, and flat $\hardnuinner$ are disfavored. Recall from \S \ref{sec:meth_detstats} that calculating $\dpss$ by integrating over the $\fstat$-statistic with zero signal present still produces a nonzero detection probability equal to the FAP, hence the floor of $\dpss \geq 10^{-3}$. 

Although the variance between $\dpss$ realizations is large, there are clear trends in how both $\dpss$ and $\dpbg$ depend on the model parameters. For $\gsmfmass$, $\mmbamp$, and $\mmbscatter$, the $\dpss$ medians behave similarly to the $\dpbg$ medians, just at lower values. The greatest difference in $\dpss$ and $\dpbg$ behavior occurs for the hardening parameters, both of which decrease $\dpbg$ by $3~\tr{dex}$, but only decrease $\dpss$ by $\lesssim\!\! 0.7~\tr{dex}$. $\gsmfnorm$ also shows significantly less impact on $\hcss$ than $\hcbg$, the $\hcss$ medians increasing only by $\sim\!\! 1.3~\tr{dex}$ compared to $\dpbg$ increasing by $\sim\!\!3~\tr{dex}$.

In \figref{fig:dpboth_wn}, there are a wide range of $\dpbg$s. However, following recent PTA results, there is considerable evidence for a GWB signal. Thus, in \figref{fig:evss_wn} we calibrate a PTA independently for each realization of each set of parameters. 
This shows how single source detection depends on each parameter, for a fixed confidence in the GWB. 
Because this `realization-calibrated' method is informed by current evidence for the GWB and allows for a more nuanced exploration of parameter space, we present this as our fiducial method for identifying where in parameter space single sources are most/least likely to be detected, with the key results being those of \figref{fig:evss_wn}. Meanwhile, \figref{fig:dpboth_wn} is useful to distinguish effects due to background calibration from direct effects on single sources.

\figref{fig:evss_wn} also includes the GW-only model, which \response{use similar `realization-calibrated' PTAs, with the resulting $\evss$ medians in dash-dotted light grey and 68\% CI shaded in light grey. Note that the $\thardf$ and $\hardnuinner$ panels include constant GW-only data because the GW-only model has no $\thardf$ or $\hardnuinner$ to vary. The rest of the GW-only results follow the same trends as the phenomenological cases but are lower by up to $1~\mathrm{dex}$.} 
This effect is similar to having a very steep $\hardnuinner$: both involve GW hardening dominating the entire PTA band.
Without phenomenological hardening speeding up the evolution, sources dwell in the PTA band longer, increasing the number of binaries contributing to the total $\hcbg$, while the individual loudest remain unaffected (see \S \ref{sec:results_phenom}). Since $\dpbg$ is calibrated to 50\%, we see the changes in $\evss$.


Given that PTA's have not yet detected a CW, very long $\thardf$ and flat $\hardnuinner$, both of which predict $\evss\!\!\sim\!\!1$, are unlikely. This is independent, but consistent with the short $\thardf$ constrained by GWB data in \citet{ng+23_astro} \response{and \citet{epta+23_physics}.} \citet{ng+23_astro} also favors flatter $\hardnuinner$, with $-0.4$ as their maximum-likelihood posterior. 
There are clear trends in the medians and 68\% CI of $\evss$ for all other model parameters, as well. Thus, CW detections can inform and constrain our astrophysical models for SMBHB populations and evolution, beyond the constraints placed by measuring the GWB amplitude. 

\paragraph{\bf{GSMF}} 
The top left panel of \figref{fig:evss_wn} shows that $\evss$ decreases smoothly from $\gsmfnorm=-3.5$ to $\gsmfnorm=-1.5$, as a result of the more significant increases in $\hcbg$ than $\hcss$ for increasing GSMF normalization.
Therefore, a larger total number of galaxies in the universe increases the likelihood of any GW detection, but disfavors CW detection for a fixed $\dpbg$.
The top right panel of \figref{fig:evss_wn} shows $\evss$ increases with $\gsmfmass$, indicating that the single source detectability increases more with GSMF characteristic mass than background detectability does. 

\paragraph{\bf{\textit{M}$_\mathbf{BH}$--\textit{M}$_\mathbf{bulge}$ Relation}}
\S \ref{sec:results_mmb} and \figref{fig:spectra_mmb} show that increasing $\mmbamp$ raises $\hcss$ slightly more than it raises $\hcbg$. Thus, the middle left panel of \figref{fig:evss_wn} shows a subtle positive trend in $\evss$ versus $\mmbamp$.
In \figref{fig:dpboth_wn} it is clear that $\mmbscatter$ affects $\dpbg$ the least of the six parameters because the impact of scatter on $\hcbg$ is minor. 
Rather, $\mmbscatter$ primarily affects the edge cases of the loudest/most massive sampled sources, i.e., those extracted as single sources. Thus, there is dramatic growth in $\dpss$ and moderate growth in $\evss$ when raising $\mmbscatter$ from $0.0$ to $0.9$. 

\paragraph{\bf{Hardening}}
\figref{fig:dpboth_wn} shows $\dpbg$ decreases with $\thardf$, except at short $\thardf$ where the loss of massive binaries due to failure to reach the PTA band becomes less significant (until a slight uptick occurs at 0.1 Gyr where decreasing distance outweighs the loss of massive sources). Single sources become nearer, but are less affected by the filtration of binaries, leading $\evss$ in the bottom left panel of \figref{fig:evss_wn} to increase to $\gtrsim\!\!1$ at long $\thardf$. 
In the bottom right panel of \figref{fig:evss_wn}, $\evss$ increases most dramatically between $-0.75 \lesssim \hardnuinner \lesssim 0$. Thus, CW detections (assuming fixed GWB confidence) become increasingly likely when phenomenological hardening processes speed up small separation ($a\lesssim \hardrchar = 10^2 ~\tr{pc}$) binary evolution. Any single source detection would make it highly unlikely that $\hardnuinner$ is steep. 

\color{black}
\subsubsection{Likely CW Source Properties: Mass, Distance, and Frequency} \label{sec:results_ds_properties}
\begin{figure}
    \centering
    \includegraphics[width=\columnwidth]{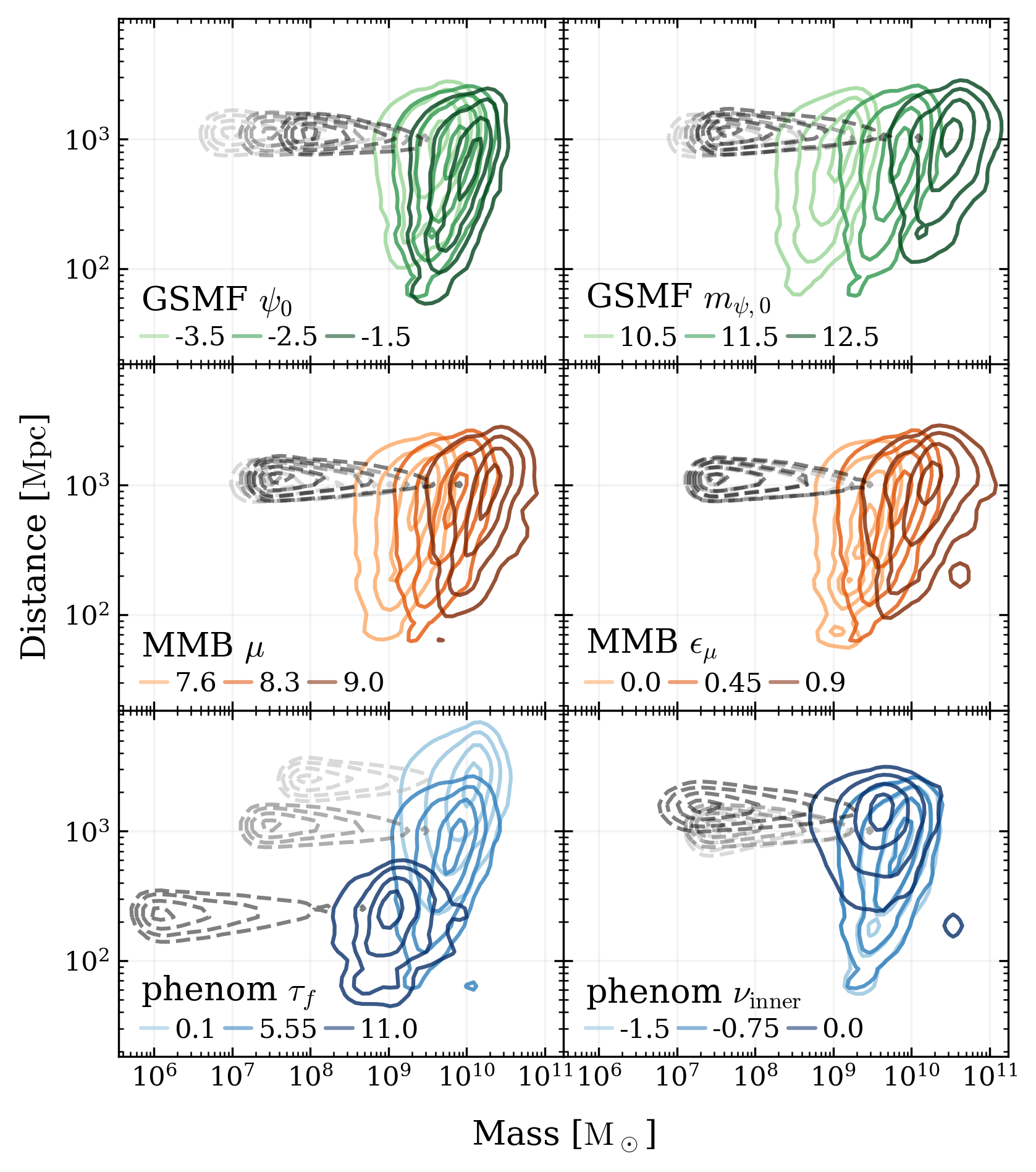}
    \caption{SNR-weighted number density of single sources' final comoving distance versus total mass. The contours represent 0.5, 1.0, 1.5, and 2.0 $\sigma$ contours for 3 variations of a single parameter, while all other parameters are fixed at their mean values. The middle shade in each plot refers to the mean-parameters model (i.e.~with the fiducial values in Table \ref{tab:varpars}). 
    The different colors correspond to the same models as in Figs. \ref{fig:spectra_gsmf}, \ref{fig:spectra_mmb}, and \ref{fig:spectra_hard}, where green, orange, and blue represent the single sources for the GSMF, $\mmbulge$ relation, and hardening parameters respectively, and shades of grey represent the $\hcbg$-weighted average values. The 10 loudest single sources at each frequency are used for the $\dpss$-weighted number densities, and all but these 10 loudest are used for the $\dpbg$-weighted number densities}
    \label{fig:snr_contours}
\end{figure}
The comoving distance $\distcom$ versus mass $M$ of the most detectable single sources are presented in \figref{fig:snr_contours}. Contour lines show the 0.5, 1.0, 1.5, and 2.0 $\sigma$ contours of the SNR-weighted number of sources in each bin of $M$ and $\distcom$, with the same panel and color convention for the six parameters as in \figref{fig:evss_wn}. The light, medium, and dark shades represent each model parameter's lowest, mean, and highest values, respectively. Solid colored lines represent the single sources, while the $\hcbg$-weighted $M$ and $\distcom$ are shown in dashed grey contours for comparison. 

Across all parameters, the 2$\sigma$ region spans masses of $\nobreak{2\times 10^8~\msol \lesssim \massss \lesssim 1.5 \times 10^{10}~\msol}$, while the mean-parameters model (with the Table \ref{tab:varpars} fiducial values) peaks at $\sim\!\!8 \times 10^{9}~\msol$. 
\response{These results are consistent with the chirp masses \citet{Rosado+2015} predict a 20-year IPTA could most likely detect, all peaking in probability between $10^{9.5}~\msol$ and $10^{10}~\msol$.} 
Long hardening times significantly decrease the peak masses, down to $\sim\!\! 10^9~\msol$ by ruling out nearby massive sources that would not have time to reach PTA-band separations. 

The mean-parameters model predicts single-source $\distcom$ peaking in SNR-weighted number at $\sim\!\!800~\tr{Mpc}$, with a 1.5$\sigma$ region spanning $70~\tr{Mpc} \lesssim \dcss \lesssim 2100~\tr{Mpc}$. 
\response{Only a small region of this parameter space overlaps with IPTA redshift predictions by \citet{Rosado+2015}, which are mostly beyond $\sim\!\!2100~\tr{Mpc}$ ($ z\!\! \gtrsim \!\! 0.56$).}
These distances are only impacted by the GSMF and $\mmbulge$ relation parameters insofar as having more massive binaries can allow more distant sources to be the most detectable, as shown by the correlation between increasing mass and greater distance in \figref{fig:snr_contours}. 
Yet, the background distances are not affected, so single sources break the degeneracy in distances for different GSMF and $\mmbulge$ parameters. The same is true of masses for varying $\gsmfmass$, $\mmbamp$, and $\mmbscatter$. 
\response{This shows how properties of single sources can break degeneracies between our model components in a way that is not possible using GWB detection alone.}

The hardening parameters, on the other hand, impact both single source and background distances by setting the time it takes binaries to reach the PTA band, as described in \S \ref{sec:results_phenom}. This is most obvious for the hardening time, where longer $\thardf$ leaves the loudest sources at peak distances as close as $\sim\!\!250~\tr{Mpc}$ by the time they are emitting nHz GWs. Similarly, when $\hardnuinner$ is very flat, the hardening timescales at separations just before the PTA band are small, meaning single sources reach these frequencies more quickly and thus at larger distances, with the SNR-weighted number peaking at $\sim\!\!1500~\tr{Mpc}$.

\begin{figure}
    \centering
    \includegraphics[width=\columnwidth]{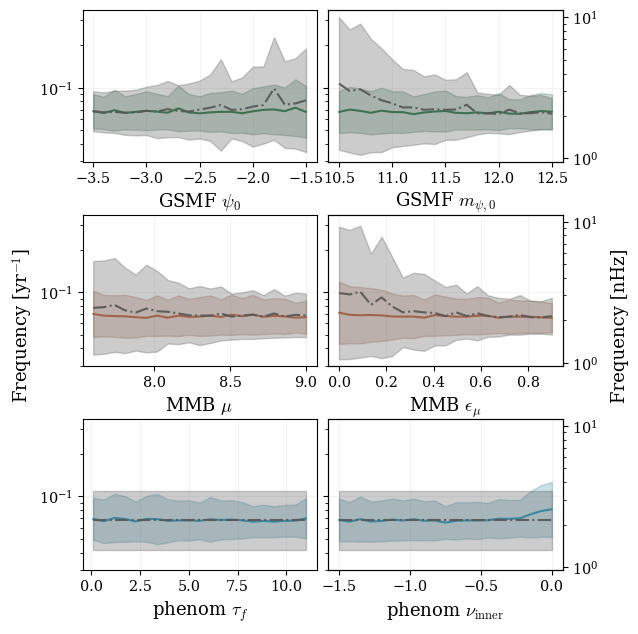}
    \caption{DP-weighted frequency of the loudest single sources, as a function of each varying parameter while the rest of the parameters are fixed. Colored regions and solid lines represent the $1\sigma$ regions and means for the phenomenological hardening model, while grey regions and grey dash-dotted lines represent the same for the GW-only hardening model.}
    \label{fig:freqs_wn}
\end{figure}

The DP-weighted average CW frequency across all single sources in all realizations $\favg$ is presented as a function of each varying parameter in \figref{fig:freqs_wn}. The color and panel convention follows that of \figref{fig:evss_wn} with the GW-only model again represented by dash-dotted grey lines. Shaded regions represent one standard deviation above and below the median in log space, calculated according to Eq.\ \ref{eq:f_std}. 
In all phenomenological cases regardless of parameters, the CW frequency most likely to be detected by our 16.03-year PTA is around $0.07~\pyr = 1.12/16.03~\tr{yr}$, except for an uptick at flat $\hardnuinner$ to $\sim\!\!0.1~\pyr$. 

For this 16.03-year PTA, $\favg$ is generally in the lowest frequency bin\response{, in agreement with similar PTA duration predictions in \citet{Rosado+2015},} because white-noise-only PTA models give a monotonic decrease in $\dpss$ versus frequency.
If $\hcss$ continues to increase with decreasing frequency, the loudest sources will likely remain in the lowest frequency bin.  
However, $\hcss$ may instead plateau at low frequencies, moving the average detection frequency to a specific value where the SS strains are maximized relative to the combined noise of the PTA and GWB \citep{Kelley+2018_ss}. 
Including red noise decreases the detection probability of the lowest frequency sources, thus moving the $\favg$ to higher frequencies. Given that pulsars typically have some intrinsic red noise \citep{ng+23_noise}, the white-noise-only $\favg$ predictions should be treated as lower limits on the predicted frequency of first CW detection. We explore the effects of varying red noise models on these predictions in \S \ref{sec:app_rednoise}.

The GW-only model mostly predicts similar frequencies to the phenomenological models, but with some increase up to $\sim\!\!0.1 ~\pyr$ at high $\gsmfnorm$ and low $\gsmfmass$, $\mmbamp$, and $\mmbscatter$ -- everywhere $\evss$ is low. 
When $\evss$ is low, the background likely becomes a more significant source of red noise, pushing the most detectable sources to higher frequencies. This also allows for more variation in the highest-DP frequency between realizations, hence the larger weighted standard deviation in the low $\evss$ regions of parameter space. 

\subsection{Anisotropy in the Gravitational Wave Background}
\label{sec:results_anisotropy}
\begin{figure}
    \centering
    \includegraphics[width=1.0 \columnwidth]{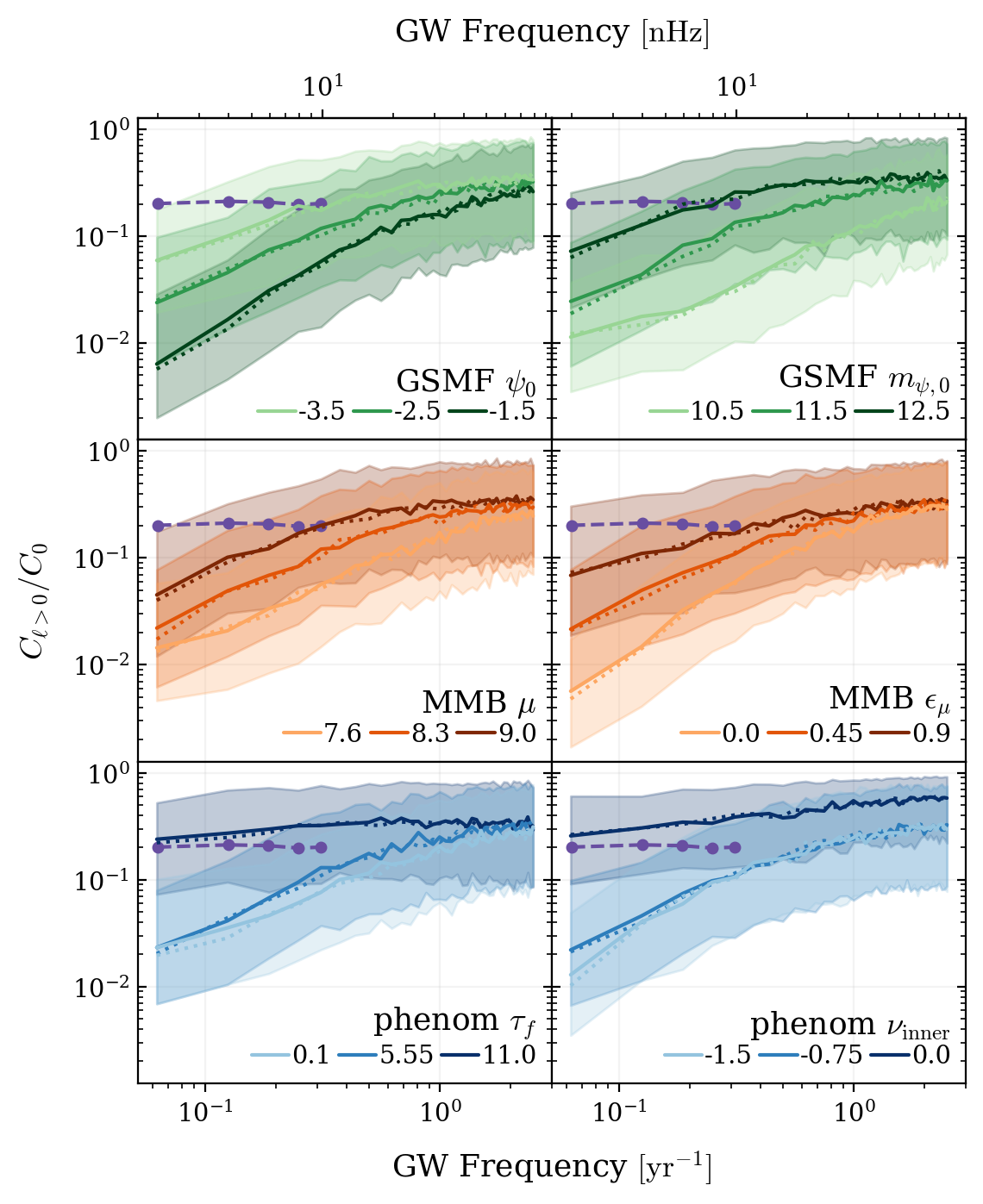} 
    \caption{Anisotropy in terms of $C_\ell/C_0$ for the first spherical harmonic mode as a function of frequency, for varying astrophysical models. Medians (solid lines) and 68\% CI (shaded) correspond to the model of the same panel and color in \figref{fig:snr_contours}, as labeled. By these methods, any $C_{\ell>0}/C_0$ is indistinguishable up to $\ell_\mathrm{max}=8$, so the $C_1/C_0$ data plotted represents any $C_{\ell>0}/C_0$ distribution. Bayesian upper limits on $C_1/C_0$ from \citet{ng+23_anis} are plotted for comparison as purple circles. 
    }
    \label{fig:anis_freq}
\end{figure}

\begin{figure}
    \centering
    \includegraphics[width=\columnwidth]{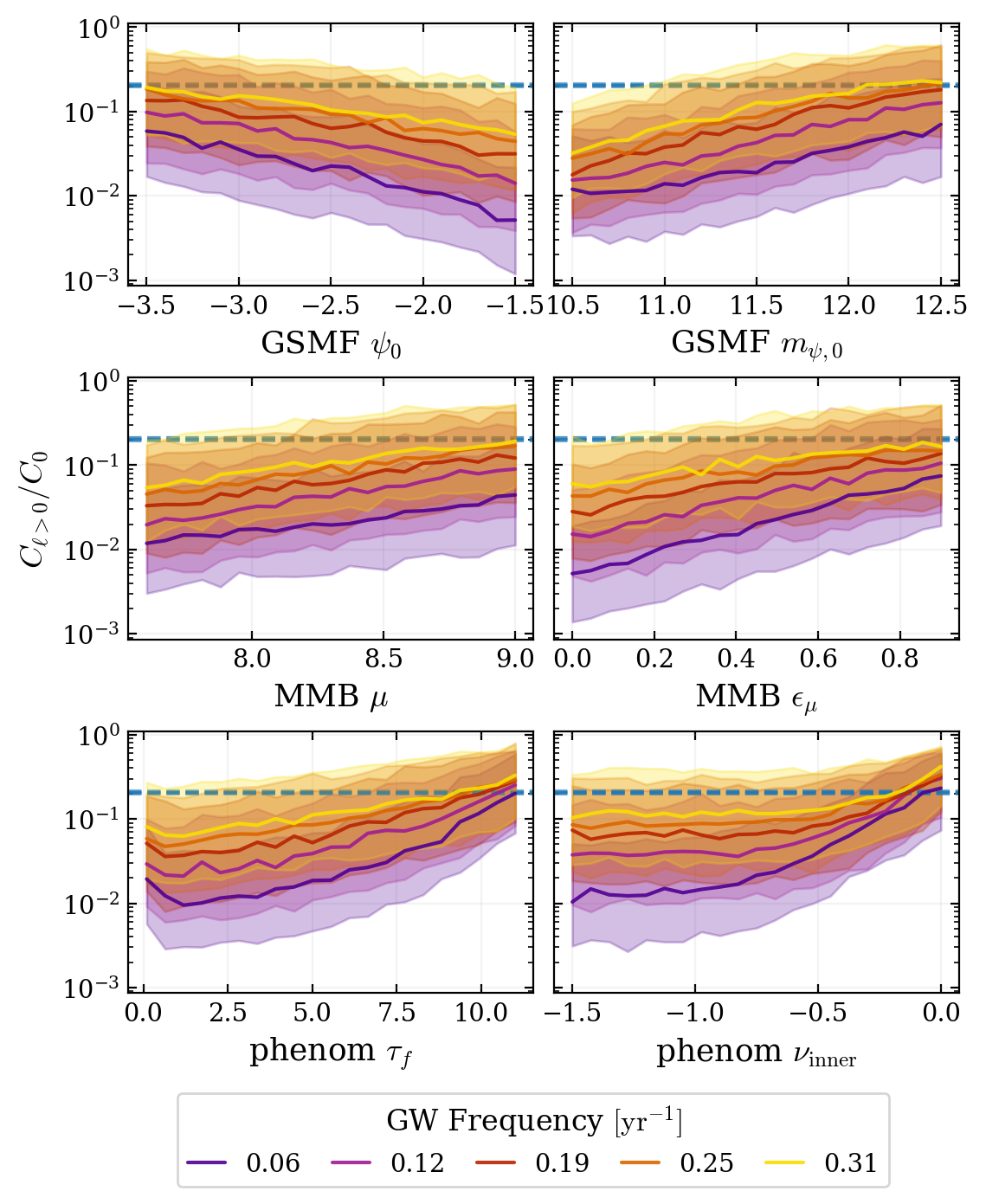}
    \caption{$C_{\ell>0}/C_0$ medians and 68\% CI for the first five frequency bins (1.98, 3.95, 5.93, 7.91, and 9.88 nHz) as a function of each varying model parameter. The data plotted uses $\ell=1$, but is indistinguishable from any other $1 \leq \ell \leq \ell_\tr{max}$. The panels correspond to the same parameters as in \figref{fig:dpboth_wn}, \figref{fig:evss_wn}, and \figref{fig:freqs_wn}. 
    These $C_\ell/C_0$s are calculated using just 10 loudest sources in each frequency bin, which \figref{fig:anis_freq} shows sufficiently reproduces the anisotropy in our model calculated using 1000 loudest sources. The \citet{ng+23_anis} upper limit on $C_1/C_0$ in the lowest frequency bin is denoted by a horizontal dashed blue line.
    }
    \label{fig:anis_vars}
\end{figure}

We calculate $C_\ell$ up to $\ell_\tr{max}=8$ for each of the models presented in \figref{fig:snr_contours}. 
The resulting $C_\ell$s are indistinguishable for each $\ell \geq 1$ of a given model, consistent with similar \texttt{holodeck} predictions marginalizing over many semi-analytic models in \citet{ng+23_anis}, \response{predictions using cosmological dark matter simulations to assign SMBHBs \citep{Taylor+Gair2013}}, and the analytic shot noise approximation for GWB anisotropy in \citet{satopolito+23}. Thus, we present $C_1/C_0$ versus frequency as a proxy for any $C_\ell/C_0$ in \figref{fig:anis_freq}, including results from the lowest, mean, and highest variation of each of the six model parameters. These are calculated using the 1000 loudest sources (solid line medians and shaded 68\% CI) and 10 loudest sources (dotted line medians) in each frequency bin. Remarkably, using the 10 loudest and the 1000 loudest sources give $C_1/C_0$ medians and standard deviations both within $\leq0.16~\tr{dex}$ of each other at any frequency, with average differences of just $\sim\!\!0.03~\tr{dex}$ and $\sim\!\!0.02~\tr{dex}$, respectively. Thus, in our models, anisotropy in the GWB is determined by $\leq10$ loudest sources in any given frequency bin. 

The medians span $\nobreak{6\times10^{-3}\lesssim C_\ell/C_0 \lesssim 2\times 10^{-1}}$ at low frequencies and $2\times 10^{-1}\lesssim C_\ell/C_0 \lesssim 6\times 10^{-1}$ at high frequencies. 
This increase in anisotropy with increasing frequency is expected because dwindling numbers of massive sources make the background $\hcbg$ drop off more than individual sources' $\hcss$, until there is hardly even a ``background" at high frequencies. 
\response{These results overlap with the $C_\ell/C_0$ of the \citet{satopolito+23} models favoring a small number of binaries with high mass and low redshift. However, the remaining three of their five models, which reproduce the same characteristic strain using a larger number of lower mass, higher redshift binaries, predict levels of anisotropy up to five orders of magnitude below ours, at low frequencies. 
Meanwhile, most of our confidence intervals overlap with the \citet{Taylor+Gair2013} calculation of $C_{\ell>0}/C_0 \!\! \sim \! \!0.1$ for a dark matter simulation-based SMBHB population, marginalized over many frequencies.}

\citet{ng+23_anis} place Bayesian upper limits of $C_1/C_0\lesssim 2\times 10^{-1}$ (circles with dashed lines in \figref{fig:anis_freq}). Most 68\% CI overlap or nearly reach these upper limits, suggesting that if the GWB is produced by SMBHBs, anisotropy may be detected in the near future, and the lack thereof could place stringent constraints on our parameter space. In fact, very long hardening time and flat hardening index predict median anisotropy levels above the current upper limits. This disfavors these corners of parameter space and supports the idea that single sources are particularly useful for constraining binary evolution.

\figref{fig:anis_vars} shows $C_1/C_0$ for the lowest five frequency bins, as a function of each varying parameter, using just ten loudest per frequency bin. In comparing \figref{fig:anis_vars} to \figref{fig:evss_wn}, it is evident that the models with the greatest $C_\ell/C_0$ correspond to those with the highest $\evss$. This is because increasing $\evss$ and increasing anisotropy both stem from cases where the loudest single sources become more dominant.
The greatest model-dependent changes in anisotropy are for long $\thardf$ and flat $\hardnuinner$. Both these scenarios produce very high $C_\ell/C_0$ ($\gtrsim 0.2$) at the lowest end of the PTA band, and then are nearly constant with frequency. 
The significant increases in $C_\ell/C_0$ at low frequencies correspond to the scenarios in \figref{fig:spectra_hard} where $\hcbg$ decreases significantly and $\hcss$ sees less change.


\section{Discussion} \label{sec:discussion}
We present the dependence of single-source detection statistics and anisotropy on astrophysical model parameters. These models include several assumptions to keep in mind. First, we assume circular orbits for all binaries. Allowing for eccentricity may move some GW energy from lower to higher frequencies, and could have different impacts on the loudest single sources versus the background -- a subject worth further investigation \citep{siwek+2023}. Another caveat is that our hardening model prescribes a fixed hardening time for all binaries, regardless of mass or redshift. This is a useful approximation to self-consistently examine how changing overall binary evolution impacts GW models without adding too many degrees of freedom, but there is no reason that these hardening times should not be mass-dependent. Thus, we suggest allowing binary lifetimes to depend on mass as a potential way to expand upon this hardening model. A third caveat to this model is that the SMBH--host galaxy relations use empirical measurements of local galaxies. These relations can be improved as more EM data is collected, particularly about more distant galaxies. The relations are based entirely on bulge mass and could be expanded by including velocity dispersion \citep{matt+2023, simon_23}.

\response{Degeneracies and covariances between model components serve as key challenges in making GW-based constraints. 
This work shows the potential for using single sources to distinguish between parameters that are degenerate in shaping the GWB. 
By considering multiple parameters, one can raise single source detection probability while maintaining the GWB amplitude by increasing the masses of binaries--whether through the GSMF $\gsmfmass$ or $\mmbulge$ $\mmbamp$--and decreasing the total number of binaries. 
The previous examples translate similarly to anisotropy. 
Fig. 6 demonstrates that the mass and distance of loud single sources can further break degeneracies, as discussed in \S \ref{sec:results_ds_properties}. 
While we defer an exploration of specific degeneracies to future work incorporating PTA data, the discrepancy in how $\evss$ and anisotropy versus the GWB amplitude vary implicitly shows that single sources can carry \textit{different} information from the GWB. Thus, we can make the strongest constraints by combining them.
}

\response{Another} challenge in making any conclusions based on single source detection statistics is the 3 dex spread of $\evss$ 95\% CIs. This is a result of the fact that CW detections depend upon the random chance of a particularly massive binary happening to be nearby. This randomness limits the precision of any single source predictions or parametric constraints by semi-analytic models. Incorporation of galaxy catalogs may allow for more narrowly constrained predictions as to what single sources could be realized in \textit{our} universe. 

EM data may also inform the level of GWB anisotropy in our universe. By placing the sources randomly and treating the background as purely isotropic, we make conservative estimates for anisotropy. However, one might predict that more SMBHBs will be emitting PTA-band GWs in regions of higher cosmic density. 
A future step would be to study possible correlations between GWB anisotropy and galaxy clustering or large-scale structure. 
On the other hand, if just a few loudest sources at distances of $\sim\!\!1000~\tr{Mpc}$ entirely determine anisotropy (as we have found), then these correlations seem less likely because the placement of individual sources is random on scales large enough to treat the universe as purely isotropic. 
Regardless, this would be an interesting hypothesis to test.  

By varying PTA noise to make each model produce 50\% $\dpbg$, we comprehensively explore the detection statistics of a wide parameter space, including models that produce low GWB amplitudes. The next step to build on this parameter space exploration is to condition our models on current measurements of the GWB amplitude. With these GWB-conditioned models, we can \response{fully explore the multi-dimensional parameter space to explicitly include covariances and} use realistic PTAs to calculate our detection statistics, as opposed to calibrating to a fixed $\dpbg$. The resulting background detection statistics will serve as a check on the GWB constraints set by \citep{ng+23_astro}. Then, we will test whether the current lack of CW detections \citep{ng+23_individuals} and upper limits on anisotropy \citep{ng+23_anis} can further constrain these model parameters. We expect that long $\thardf$ and flat $\hardnuinner$ will be most easily constrained by single source detection statistics and anisotropy, given that these are the regions of parameter space where both $\evss$ and $C_\ell/C_0$ are highest and $\evss$ has the lowest variance.



\color{black}






\section{Conclusions} \label{sec:conclusions}
In this work, we develop an approach for modeling CWs distinguishable from a background of SMBHBs, their sources' properties, and their corresponding GWB anisotropy. We develop a detection statistics pipeline that calibrates a simulated PTA to a 50\% probability of detecting the background and calculates the expected number of single-source detections under those settings. Our primary conclusions are the following:
\begin{enumerate}
    \item GW anisotropy and CW detections (or lack thereof) convey specific information about the astrophysics governing galaxy population, their SMBHBs, and binary evolution.
    This anisotropy and CW information can break model degeneracies and allow for much more stringent constraints than possible with the GWB alone.

    \item CWs are increasingly likely to be observed for low GSMF normalization $\gsmfnorm$, high GSMF characteristic mass $\gsmfmass$, high $\mmbulge$ mass normalization $\mmbamp$ and intrinsic scatter $\mmbscatter$, long hardening time $\thardf$, and flat small-separation hardening index $\hardnuinner$.

    \item Anisotropy in the GWB, represented by the angular power spectrum $C_\ell$ over multipole modes $\ell$, is determined in our models by $\leq\!\! 10$ loudest sources at each frequency and is the same for any $1 \leq \ell \leq \ell_\tr{max}$. Models vary in their low-frequency predictions for $C_{\ell>0}/C_0$  from $\nobreak{ \sim\!\! 5\times 10^{-3}}$ to $\nobreak{\sim\!\!2\times10^{-1}}$ and converge to $\nobreak{\sim\!\! 3\times 10^{-1}}$ at high frequencies. 
        
    \item Models with greater single-source detection probability tend to have higher anisotropy, often exceeding current upper limits of $C_1/C_0\lesssim 0.2$ at low frequencies. Thus, not detecting anisotropy could strongly constrain our parameter space. 
    
    \item \response{Of all our model components, binary evolution shows the greatest promise for constraints using single sources and anisotropy.} Long hardening time and flat $\hardnuinner$ give the greatest probability of CW detection for a fixed GWB confidence and high anisotropy even at low frequencies.
    
    \item The most detectable single sources are found in the lowest frequency bin for a 16.03-year PTA with 
    masses ranging from $\nobreak{\sim\!\!10^9~\msol}$ to $\nobreak{\sim\!\!3 \times 10^{10}~\msol}$ 
    and final comoving distances ranging from $\nobreak{\sim\!\!250~\tr{Mpc}}$ to $\nobreak{\sim\!\! 2500~\tr{Mpc}}$. 
    The most detectable frequency has little dependence on the model but increases with greater pulsar red noise.

    \item
    Single source masses generally increase with increasing $\gsmfnorm$, $\gsmfmass$, $\mmbamp$, $\mmbscatter$, and decreasing hardening time.
    Only the hardening parameters have a demonstrable impact on these sources' final comoving distances, with longer $\thardf$ and steep $\hardnuinner$ resulting in the closest sources.

\end{enumerate}


\section{Acknowledgments}
The authors thank Jeff Hazboun, Jeremy Baier, Bence B\'ecsy, and Neil Cornish for helpful discussions throughout the development of this paper, particularly regarding CW detection statistics. We also thank Nihan Pol for insight into the anisotropy methods and interpretation. Finally, we thank the
anonymous referee for thoughtful and constructive feedback. The work of AM and AL was supported by the Deutsche Forschungsgemeinschaft under Germany’s Excellence Strategy - EXC 2121 Quantum Universe - 390833306.

\software{
    astropy \citep{astropy},
    cython \citep{cython2011},
    hasasia \citep{Hazboun2019Hasasia},
    healpy \citep{healpy},
    HEALpix \citep{healpix}
    holodeck \citep{holodeck},
    jupyter \citep{jupyter2016},
    kalepy \citep{kalepy},
    matplotlib \citep{matplotlib2007},
    numpy \citep{numpy2011},
    scipy \citep{2020SciPy-NMeth},
}

%






\appendix

\section{PTA Noise Models}
\label{sec:app_rednoise}

\begin{figure}
    \centering
    \includegraphics[width=1.0\columnwidth]{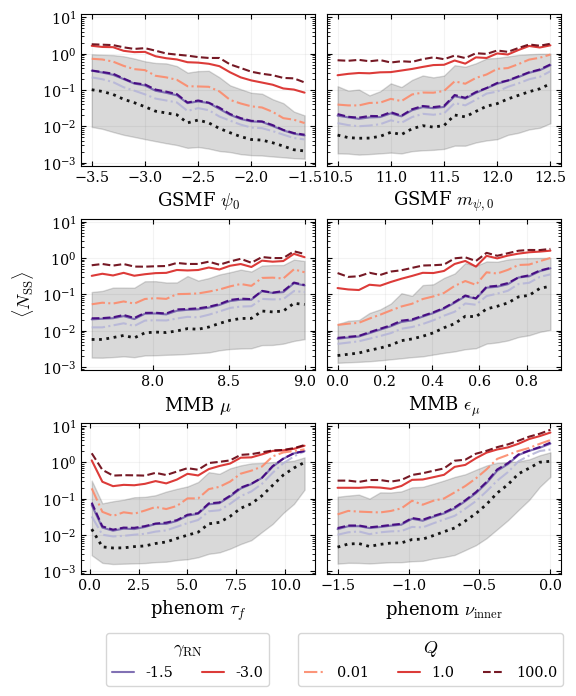}
    \caption{Expected number of single-source detections when the background detection probability is calibrated to 50\% by varying the pulsar noise levels. The noise is set by a fixed ratio $Q = S_\tr{RN}(f_\tr{ref})/S_\tr{WN}(f_\tr{ref})$ between white noise and red noise at reference frequency $f_\tr{ref}=1~\pyr$, and fixed red noise index of $\gamma_\tr{RN}=-1.5$ (purple), or $-3.0$ (red). Noise ratios of 0.01, 1.0, and 100.0 are represented by dash-dotted, solid, and dashed lines, respectively, and the 68\% CI of the white noise-only results are shaded. }
    \label{fig:evss_rn}
\end{figure}

\begin{figure}
    \centering
    \includegraphics[width=1.0\columnwidth]{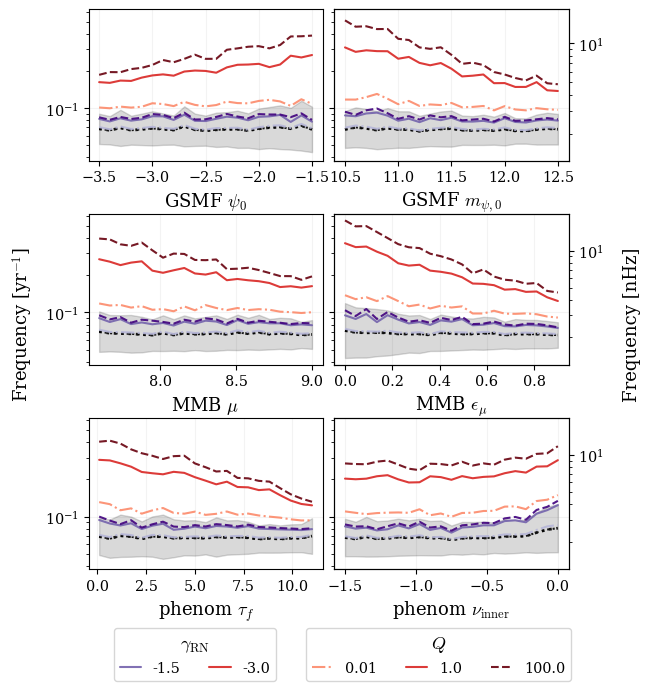}
    \caption{$\dpssi$-weighted frequency of the loudest single sources as a function of varying model parameters. The color and line style conventions follow those of \figref{fig:evss_rn}, with shaded regions showing the weighted standard deviation of the white noise-only results.}
    \label{fig:favg_rn}
\end{figure}

The primary work in this paper uses a white-noise-only simulated PTA. We also consider the impact of red noise models, parameterized by an amplitude $\redamp$ at reference frequency $f_\tr{ref} = 1~\tr{yr}^{-1}$ and power-law index $\redgamma$,
\begin{equation}
    S_\tr{RN} = \frac{\redamp^2}{12\pi^2}\scale[\redgamma]{f}{f_\tr{ref}} f_\tr{ref}^{-3},
\end{equation}
on our detection probabilities and $\dpss$-weighted average frequencies. The red-noise PTAs are calibrated by fixing $\gamma_\tr{red}$ and the ratio $Q$ of red noise to white noise 
\begin{equation}
    Q \equiv \frac{S_\tr{RN}(f_\tr{ref})}{S_\tr{WN} }
\end{equation}
while allowing the total noise amplitude to vary. 

\figref{fig:evss_rn} shows the resulting $\evss$ as a function of the six varying model parameters for the fiducial white noise only model (black), red-noise with spectral index $\redgamma=-1.5$ (purple), and red-noise with spectral index $\redgamma=-3.0$ (red). For the red-noise models we include ratios of $Q=0.01$ (dash-dotted), $Q=1.0$ (solid), and $Q=100$ (dashed). The addition of red noise generally raises the relative single source detection probability, because it makes the GWB less distinguishable from the noise, such that the total noise calibration must be lower for a 50\% $\dpbg$. In fact, when steep ($\redgamma=-3.0$) red noise dominates ($Q=100$), the median $\evss$ is $\gtrsim0.1$ for all model variations except the highest $\gsmfnorm$s. The increase in $\evss$ from previously low regions results in an overall flattening in parametric dependence, but maintains the sign of the derivative, i.e.~whether $\evss$ is increasing or decreasing with each parameter.

\figref{fig:favg_rn} shows that adding red noise also raises the most detectable CW frequency because lower frequency sources are drowned out. Moderate red noise ($\redgamma=-1.5$, purple) maintains the lack of dependence on model parameters seen in the white noise cases. However, adding steep red noise ($\redgamma=-3.0$, red) with a ratio of $Q\gtrsim 1$ creates a dependence of $\favg$ on each varying parameter. For all but $\hardnuinner$, this $\favg$ dependence tends to follow the opposite trend of $\evss$. When $\evss$ is low, there is greater noise from the PTA calibration, especially at low frequencies,  pushing $\favg$ higher. 



\bibliography{refs}{}
\bibliographystyle{aasjournal}



\end{document}